\documentclass{aa}  

\usepackage{tabularx}
\usepackage[]{hyperref}
\usepackage{rotating}
\usepackage{acronym}
\usepackage{booktabs}
\usepackage{units} 
\usepackage{glossaries}
\usepackage{subfigure}
\usepackage{graphicx}
\usepackage{rotating}
\usepackage{multirow}
\usepackage{orcidlink}
\usepackage{url}
\usepackage{caption}
\usepackage[utf8]{inputenc}

\acrodef{EBL}[EBL]{Extragalactic Background Light}
\acrodef{Fermi-LAT}[LAT]{Large Area Telescope}
\acrodef{FGST}[FGST]{Fermi Gamma-Ray Space Telescope}
\acrodef{PSD}[PSD]{Power Spectral Density}
\acrodef{SSC}[SSC]{Synchrotron Self-Compton}
\acrodef{SED}[SED]{spectral energy distribution}
\acrodef{EC}[EC]{External Compton}
\acrodef{IC}[IC]{Inverse Compton}
\acrodef{FSRQ}[FSRQ]{Flat Spectrum Radio Quasar}
\acrodef{CD}[CD]{Compton Dominance}
\acrodef{VLBI}[VLBI]{Very Long Baseline Interferometry}
\acrodef{VLBA}[VLBA]{Very Long Baseline Array}
\acrodef{MOJAVE}[MOJAVE]{Monitoring Of Jets in Active galactic nuclei with VLBA Experiments}
\acrodef{VLBA-BU-BLAZAR}[VLBA-BU-BLAZAR]{VLBA-BU Blazar Monitoring}
\acrodef{FSRQ}[FSRQ]{Flat Spectrum Radio Quasar}
\acrodef{BLR}[BLR]{broad-line region}
\acrodef{Fermi}[\textit{Fermi}]{\textit{Fermi Gamma-ray Space Telescope}}
\acrodef{Swift}[\textit{Swift}]{\textit{Neil Gehrels Swift Observatory}}
\acrodef{XRT}[XRT]{X-Ray Telescope}
\acrodef{UVOT}[UVOT]{Ultraviolet/Optical Telescope}
\acrodef{CRTS}[CRTS]{Catalina Real-Time Transient Survey}
\acrodef{KAIT}[KAIT]{Katzman Automatic Imaging Telescope}
\acrodef{Tuorla}[Tuorla]{Tuorla Observatory}
\acrodef{Atlas}[ATLAS]{Irsa's Image and Spectrum Server}
\acrodef{ZTF}[ZTF]{Zwicky Transient Facility}
\acrodef{F-GAMMA}[F-GAMMA]{FERMI-GST AGN Multi-frequency Monitoring Alliance}
\acrodef{MOJAVE}[MOJAVE]{Monitoring Of Jets in Active galactic nuclei with VLBA Experiments}
\acrodef{SMA}[SMA]{Submillimeter Array}
\acrodef{VLBA-BU-BLAZAR}[VLBA-BU-BLAZAR]{VLBA-BU-BLAZAR}
\acrodef{MWL}[MWL]{multiwavelength}
\acrodef{4FGL}[4FGL]{fourth \textit{Fermi} Large Area Telescope}
\acrodef{2FGL}[2FGL]{second \textit{Fermi} Large Area Telescope}
\acrodef{3FGL}[3FGL]{third \textit{Fermi} Large Area Telescope}
\acrodef{GRB}[GRB]{gamma-ray burst}
\acrodef{BAT}[BAT]{Burst Alert Telescope}
\acrodef{LST}[LST]{Large-Sized Telescope}
\acrodef{AGN}[AGN]{Active Galactic Nuclei}
\acrodef{HE}[HE]{high energy}
\acrodef{VHE}[VHE]{very high energy}
\acrodef{CTAO}[CTAO]{Cherenkov Telescope Array Observatory}
\acrodef{DT}[DT]{dusty torus}

\newcommand{\flAGNs}{Active Galactic Nuclei (AGN) }
\newcommand{\op}{OP 313}

\begin{document}
\authorrunning{C. Bartolini}
    \title{A long-term multiwavelength study of the flat spectrum radio quasar OP 313}
    \author{C. Bartolini\, \orcidlink{0000-0001-7233-9546} \inst{1, 2, 3} \thanks{Corresponding authors: C.~Bartolini, E.~Lindfors, e-mail:   {\fontfamily{cmtt}\selectfont chiara.bartolini-1@unitn.it, elilin@utu.fi}}\, E. Lindfors\, \orcidlink{0000-0002-9155-6199} \inst{4}\footnotemark[1], A. Tramacere\, \orcidlink{0000-0002-8186-3793} \inst{5}, M. Giroletti\, \orcidlink{0000-0002-8657-8852} \inst{6}, D. Cerasole\, \orcidlink{0000-0003-2033-756X} \inst{2, 3}, I. Agudo\, \orcidlink{0000-0002-3777-6182
} \inst{7}, E. Angelakis\, \orcidlink{0000-0001-7327-5441
} \inst{8}, E. Bissaldi\, \orcidlink{0000-0001-9935-8106} \inst{2, 3}, F. Casaburo\, \orcidlink{0000-0002-2260-9322} \inst{9, 10, 11},  F. D'Ammando\, \orcidlink{0000-0001-7618-7527
} \inst{6},  L. Di Venere\, \orcidlink{0000-0003-0703-824X} \inst{3}, 
V. Fallah Ramazani\, \orcidlink{0000-0001-8991-7744} \inst{12}, F. Giacchino\, \orcidlink{0000-0002-0247-6884} \inst{13, 9, 10},   F. Giordano\, \inst{2, 3}, M. Gurwell\, \orcidlink{0000-0003-0685-3621} \inst{14}, J. Jormanainen\, \orcidlink{0000-0003-4519-7751} \inst{12, 4}, S. Jorstad\, \orcidlink{0000-0001-6158-1708} \inst{15}, G. Keating\, \orcidlink{0000-0002-3490-146X} \inst{14}, P. M. Kouch\, \orcidlink{0000-0002-9328-2750} \inst{12, 4, 17},  A. Kraus\, \orcidlink{0000-0002-4184-9372} \inst{16}, A. L\"ahteenm\"aki\, \orcidlink{0000-0002-0393-0647} \inst{18, 17}, S. Loporchio\, \orcidlink{0000-0003-4457-5431} \inst{3}, N. Marchili\, \orcidlink{0000-0002-5523-7588} \inst{6}, A. Marscher\, \orcidlink{0000-0001-7396-3332} \inst{15}, I. Myserlis\, \orcidlink{0000-0003-3025-9497} \inst{19, 16}, R. Rao \orcidlink{0000-0002-1407-7944} \inst{14}, S. Righini\, \orcidlink{0000-0001-7332-5138} \inst{6}, M. Tornikoski\, \orcidlink{0000-0003-1249-6026} \inst{17}}
    \institute{Università di Trento, via Sommarive, 14 - I-38123 Trento (Italy) \and Dipartimento di Fisica “M. Merlin” dell’Università e del Politecnico di Bari, via Amendola 173, I-70126 Bari, Italy \and Istituto Nazionale di Fisica Nucleare, Sezione di Bari, I-70126 Bari, Italy \and Department of Physics and Astronomy,
University of Turku - FI-20014 Turku, Finland \and  Department of Astronomy, University of Geneva, Chemin Pegasi 51, 1290, Versoix, Switzerland \and Istituto di Radioastronomia - INAF, Via Piero Gobetti 101, I-40129 Bologna, Italy \and Instituto de Astrof\'{i}sica de Andaluc\'{i}a-CSIC, Glorieta de la Astronomia, E-18008, Granada, Spain \and Section of Astrophysics, Astronomy \& Mechanics - Department of Physics, National and Kapodistrian University of Athens - Panepistimiopolis Zografos 15784, Greece \and  Istituto Nazionale di Fisica Nucleare, Sezione di Roma “Tor Vergata”, I-00133 Roma, Italy \and ASI Space Science Data Center, via del Politecnico, I-00133, Roma, Italy \and Sapienza Università di Roma- Dipartimento di Fisica, P.le A. Moro, 00185, Roma, Italy \and  Finnish Centre for Astronomy with ESO, University of Turku, Finland \and Department of Fundamental Physics, University of
Salamanca, Plaza de la Merced s/n, E-37008 Salamanca, Spain \and Center for Astrophysics | Harvard \& Smithsonian, 60 Garden Street, Cambridge, MA 02138, USA \and Institute for Astrophysical Research, Boston University, 725 Commonwealth Ave., Boston, MA 02215, USA \and Max-Planck-Institut f\"ur Radioastronomie - Auf dem H\"ugel 69, D-53121 Bonn, Germany \and Aalto University Metsähovi Radio Observatory - Metsähovintie 114, 02540 Kylmälä, Finland \and Aalto University Department of Electronics and Nanoengineering, P.O. BOX 15490, FI-00076 AALTO, Finland \and Institut de Radiostronomie Millim\'{e}trique - Avenida Divina Pastora 7, Local 20, E–18012 Granada, Spain }

\date{Received *; accepted *}
 
\abstract
{The Flat Spectrum Radio Quasar \op\, is a high-redshift ($z=0.997$) blazar that entered an intense $\gamma$-ray active phase from November 2023 to March 2024, as observed by the \ac{Fermi-LAT} on board the \textit{Fermi Gamma-ray Space Telescope}.}
{We present a multiwavelength analysis covering 15 years of data, from August 2008 to March 2024, to contextualize this period of extreme $\gamma$-ray activity within the long-term emission of the source.} 
{We analyzed a long-term, comprehensive, multiwavelength dataset from different facilities and projects from radio to $\gamma$ rays. We identified the 7 most intense $\gamma$-ray flaring periods and performed a kinematic analysis of Very Long Baseline Array (VLBA) data to determine whether new jet components emerged before or during these flares. For 2 of these flaring periods, we performed the modeling of the spectral energy distribution (SED).}
{The VLBA-BU-BLAZAR and MOJAVE datasets reveal a new jet component appearing in both visibility datasets prior to the onset of one of the strongest $\gamma$-ray flares. By comparing the timing of the VLBA-BU-BLAZAR knots ejection with the $\gamma$-ray flaring periods, we constrained the setup of the SED modeling. We also found that the first $\gamma$-ray flaring period is less Compton-dominated than the others.} 
{Our results suggest that the recent activity of \op\, is triggered by new jet components emerging from the core and interacting with a standing shock. The $\gamma$-ray emission likely arises from dusty torus photons upscattered via \ac{IC} by relativistic jet electrons. The SED modeling indicates that this component is less dominant during the first $\gamma$-ray flaring period than the later ones.}
\maketitle

\section{Introduction} \label{sec:intro}

Relativistic jets launched by \flAGNs are among the most energetic phenomena in the Universe and play a fundamental role in regulating galaxy evolution via feedback processes \citep{blandford_relativistic_2019}. Blazars are a subclass of AGNs with their relativistic jet directed towards us. They are classified into two categories: \acp{FSRQ} and BL Lacs (BLL). \acp{FSRQ} exhibit strong and broad optical emission lines and high bolometric luminosities \citep{sambruna_spectral_1996}. They often display signs of thermal emission related to the accretion disk in their optical/UV spectra \citep{smith_optical_1986}.

On the other hand, BL Lacs show weak optical lines and lower bolometric luminosities than FSRQs. Both categories exhibit extremely variable emission across the entire electromagnetic spectrum, on timescales ranging from months to minutes in the GeV and TeV bands \citep{aharonian_exceptional_2007, ackermann_minute-timescale_2016}. The physical mechanisms driving their variability remain unclear. Emission variability at different timescales may arise from distinct processes occurring in different regions within the AGN. The detection of high-energy (HE, $0.1-100\, \mathrm{GeV}$) and very high-energy (VHE, $\geq100\, \mathrm{GeV}$) emission from FSRQs suggests an origin further out in the jet, where the absorption of $\gamma$ rays by external radiation fields, such as from the \ac{BLR}, is reduced \citep[e.g., ][]{donea_radiation_2003}.

The \ac{SED} of blazars shows two peaks. The low-energy bump extends from the radio to the soft X-ray band and is interpreted as synchrotron emission produced by highly relativistic electrons within the jet. The high-energy bump extends up to $\gamma$-ray energies and its origin is still debated. In leptonic models, this bump originates from relativistic electrons in the jet upscattering low-energy photons through the inverse Compton process. Primarily, the low-energy photon field is provided by the synchrotron radiation within the jet, in a process known as Synchrotron Self-Compton \citep[SSC, see, e.g.,][]{inoue_electron_1996}. In addition, low-energy seed photons from regions external to the jet, such as the accretion disk, the \ac{BLR}, and the \ac{DT}, can be upscattered via External inverse Compton \citep[EC, see][]{dermer_model_1993, dermer_gamma-ray_2009}. In lepto-hadronic models, the interactions between relativistic protons and low-energy photon fields produce high-energy secondary particles, including $\gamma$ rays and neutrinos \citep[e.g.,][]{kelner_energy_2008, rodrigues_multiwavelength_2021}. 

Indeed, blazars play a crucial role in multifrequency and multimessenger astrophysics, and some of them have been proposed as astrophysical neutrino sources \citep[][]{icecube_collaboration_multimessenger_2018, garrappa_investigation_2019, icecube_collaboration_evidence_2022}. Nonetheless, both leptonic and lepto-hadronic models are usually capable of reproducing the observed emission from blazars \citep[e.g.,][]{mannheim_gamma-ray_1992, aharonian_tev_2000, mucke_proton_2001, mucke_bl_2003, bottcher_leptonic_2013}.

\op, also known as B2 1308+326, is a \ac{FSRQ} at a redshift of $z=0.997$ \citep{schneider_sloan_2010} and coordinates $\text{R.A.} = 197.619433$ deg and $\text{Dec} = 32.345495$ deg \citep{xu_structure_2019}. 
Several observational campaigns targeting this source have been performed in the past due to its variability and uncertain classification \citep[e.g.,][]{britzen_swirling_2017}. In \cite{stickel_1991apj374431s_1991}, \op\, was classified as a BLL due to its nearly featureless optical spectrum \citep{wills_spectroscopy_1979}, high linear polarization in the optical band \citep[$> 3\%$,][]{jannuzi_optical_1993}, and extreme optical variability \citep{angel_optical_1980}. Nevertheless, based on \ac{VLBI} measurements, \cite{gabuzda_is_1993} reported polarized emission from the inner jet of OP 313 with a position angle orthogonal to the jet direction and consequently identified the object as a \ac{FSRQ}.

\cite{watson_asca_2000} observed \op\, from radio to X rays and suggested that it may be either a transitional blazar or a gravitationally microlensed quasar.

The Large Area Telescope (LAT) onboard the \textit{Fermi Gamma-Ray Space Telescope} detected flaring activity from \op\, for the first time in April 2014, with an average $\gamma$-ray flux above $100\, \mathrm{MeV}$ of $(1.1 \pm 0.2) \times 10^{-6}\, \mathrm{ph/cm^{2} s^{1}}$ \citep{buson_atel_2014}. Two ATels were published in 2021 and 2022 \citep{mangal_atel_2021, garrappa_atel_2022}, reporting periods of re-enhanced activity from the source. 

\op\, entered the phase of its highest activity in HE $\gamma$ rays, as detected by the \textit{Fermi}-LAT, on the 22nd of November, 2023.

The first peak of activity was reached on the 24th of November 2023 with an average $\gamma$-ray flux at energies above $100\, \mathrm{MeV}$ of $(1.8\pm0.2)\times 10^{-6}\, \mathrm{photons/ cm^{2}s}$, approximately 40 times larger than the flux reported in the \ac{4FGL} catalog \citep[$2.4\times 10^{-9}\, \mathrm{photons/cm^{2}s}$ from 0.1-100 $\mathrm{GeV}$][]{abdollahi_fermi_2020, abdollahi_incremental_2022}. Additionally, the photon index was significantly harder ($1.80\pm0.06$) than the \ac{4FGL} catalog value of $2.34\pm0.02$. This flare led the \textit{Fermi}-LAT Collaboration to publish an ATel on the 2023 high-state activity of the source \citep{bartolini_atel_2023}. 
During this high state, VHE $\gamma$-ray emission from \op\, has been detected for the first time in December 2023 by the first prototype of the Large-Sized Telescope (LST-1) of the \ac{CTAO}, located on the island of La Palma \citep{cortina_atel_2023}.

This detection made \op\, the most distant quasar detected at VHE $\gamma$ rays so far.
Another major flare from this source was detected on the 27th of February, 2024, leading to the highest flux ever detected by the \ac{Fermi-LAT} for OP 313: $(3.1\pm 0.4) \times 10^{-6}\, \mathrm{photons/cm^{2}s}$, 60 times larger than the flux reported in the 4FGL catalog, with a photon index of $1.81\pm0.09$. This detection was reported in another ATel by the \ac{Fermi-LAT} collaboration \citep{bartolini_atel_2024}.

Given that OP\,313 exhibited persistent high-state activity since 2022, following more than a decade of quiescence, we analyzed 15 years of data from radio to HE $\gamma$ rays to unveil the mechanisms responsible for the observed flares. We note that the detailed study of the VHE emission period is not included in this paper, but it will be investigated in an upcoming publication by the MAGIC and LST Collaborations. 

In Section \ref{sec:multi}, we present the multiwavelength dataset the analyses performed in each energy band.
In Section \ref{results}, we present the 15-year-long multiwavelength lightcurve and identify the most relevant $\gamma$-ray flares covering the period from January 2022 to March 2024. In addition, we present and model the broadband SEDs collected during 2 of the highest flaring activity periods.
Furthermore, we investigate the jet kinematics using an extensive \ac{VLBA} dataset at 15$\,\mathrm{GHz}$ and 43$\,\mathrm{GHz}$, as well as the correlation between the optical and $\gamma$-ray lightcurves.
We summarize the main results and present our conclusions in Section \ref{Conclusion}.

\section{Multiwavelength Data} \label{sec:multi}
This section presents the \ac{MWL} dataset employed for the investigation of \op\, across the following energy bands: the radio band from $2.64\, \mathrm{GHz}$ to $353\, \mathrm{GHz}$, the optical/UV bands from $170$ to $690\, \mathrm{nm}$, the X-ray band from $0.3$ to $10\, \mathrm{keV}$ and the $\gamma$-ray one from $20$ to $300\, \mathrm{GeV}$. Table \ref{tab:instruments} summarizes the facilities whose data have been included in this work, as well as the time span of the observations.

\begin{table*}
    \centering
    \caption{Instruments whose datasets have been included for the long-term \ac{MWL} analysis of OP 313.} 
    \label{tab:instruments}
    \begin{tabular}{cccc}
    \toprule
        Instrument & Sensitivity range & MJD start & MJD stop \\
    \midrule
    \textit{Fermi}-LAT & $20\, \mathrm{
    MeV}-300\, \mathrm{GeV}$ & 54682 & 60378\\
    \textit{Swift}-XRT & $0.2-10\, \mathrm{keV}$ & 54698 & 60378\\
    \multirow{2}{*}{\textit{Swift}-UVOT} & \textit{v, b, u, w1, m2, w2} & \multirow{2}{*}{54698} & \multirow{2}{*}{60378}\\
    & $170-650\, \mathrm{nm}$ & & \\
    \multirow{2}{*}{ZTF} & g, r, i & \multirow{2}{*}{58202} & \multirow{2}{*}{60162}\\
    & $464, 658, 806\, \unit{nm}$ & & \\
    \multirow{2}{*}{CRTS} & V & \multirow{2}{*}{54807} & \multirow{2}{*}{57505} \\
    & $551\, \unit{nm}$ & & \\
    \multirow{2}{*}{ATLAS} & o, c & \multirow{2}{*}{57362} & \multirow{2}{*}{60375}\\
    & $534, 690\, \unit{nm}$ & & \\
    \multirow{2}{*}{KAIT} & R & \multirow{2}{*}{55938} & \multirow{2}{*}{58966} \\
    & $658\, \unit{nm}$ & & \\
    Tuorla & R & 55733 & 60425 \\
    VLBA-BU-BLAZAR & $7\, \mathrm{mm}$ & 54922 & 60391 \\
    Mets$\ddot{a}$hovi & $8\, \mathrm{mm}$ & 54704 & 60132 \\
    MOJAVE  & $2\, \mathrm{cm}$ & 54838 & 60407 \\
    SMA and SMAPOL &  $0.9$, $1.1$, $1.3\, \mathrm{mm}$ & 54722 & 60379.5 \\
    \multirow{2}{*}{QUIVER} & $0.7$, $0.8$, $1.4$, $2.1\, \mathrm{cm}$ & \multirow{2}{*}{60261} & \multirow{2}{*}{60399} \\
    & $2.9$, $6.2$, $11.6\, \mathrm{cm}$ & & \\
    \multirow{2}{*}{F-GAMMA} & $2.9, 3.6, 6.2, 11.4\, \mathrm{cm}$& \multirow{2}{*}{54700} & \multirow{2}{*}{56809} \\
    & $0.7, 0.9, 1.3, 2.1,\, \mathrm{cm}$ & & \\
    \bottomrule
    \end{tabular}
    \tablefoot{The sensitivity range and corresponding time span are reported. For the optical/UV photometric filters included, the effective midpoint wavelengths are reported. MJD 54682 (MJD 60378) corresponds to 4th of August, 2008 (9th of March, 2024).}
\end{table*}

\subsection{Fermi-LAT data} \label{sec:fermi}
The \ac{Fermi-LAT} is an imaging, wide field-of-view (FoV), pair-conversion $\gamma$-ray telescope covering an energy range from below $20\, \unit{MeV}$ to more than $300\, \unit{GeV}$ \citep{atwood_large_2009}. 

We analyzed 15 years of Pass 8 data \citep{atwood_pass_2013, bruel_fermi-lat_2018}  collected by the \ac{Fermi-LAT} using the standard \ac{Fermi-LAT} \textit{ScienceTools}\footnote{\url{https://fermi.gsfc.nasa.gov/ssc/data/analysis/}} v2.2 and the \texttt{fermipy} \citep{wood_fermipy_2018} v1.1 Python package. We used the P8R3$\textunderscore $SOURCE$\textunderscore$V3 instrument response functions, the Third Data Release 4FGL-DR3 \citep{abdollahi_incremental_2022}, and the Galactic and the isotropic diffuse gamma-ray emission models provided by the standard templates $gll\textunderscore iem\textunderscore v07.fits$ and $iso\textunderscore P8R3\textunderscore SOURCE\textunderscore V3\textunderscore v1.txt$ \footnote{\url{https://fermi.gsfc.nasa.gov/ssc/data/access/lat/BackgroundModels.html}}. We analyzed photon events with reconstructed energy between $100\, \unit{MeV}$ and $300\, \unit{GeV}$, selecting a region of interest (ROI) of $15 \times 15\, \unit{deg}^2$ centered on \op\, and applying a standard quality cut ($‘\texttt{DATA}\textunderscore \texttt{QUAL>0} \quad\texttt{\&\&} \quad \texttt{LAT}\textunderscore \texttt{CONFIG==1}’$). Furthermore, a zenith angle cut of $90\, \unit{deg}$ was used to reduce the contamination from the Earth limb. 

We modeled the spectrum of \op\, with a log-parabola function\footnote{We employed the \texttt{LogParabola} model as defined in \url{https://fermi.gsfc.nasa.gov/ssc/data/analysis/scitools/source_models.html}.}.
We performed a maximum-likelihood analysis and defined the test statistic (TS) of \op\, as $TS =2(\log L_{1}-\log L_{0})$, $L_{1}$ and $L_{0}$ being the likelihood of the data for a model with and without a point source located at the position of OP 313, respectively.
In the fit, we left free to vary the parameters of the \op\, spectral model, as well as the normalizations of all sources with $TS \geq 25$ located within 5$\, \mathrm{deg}$ from \op, and the normalizations of the Galactic and isotropic diffuse components.

\begin{figure*}
    \centering
    \includegraphics[width=1.0\linewidth]{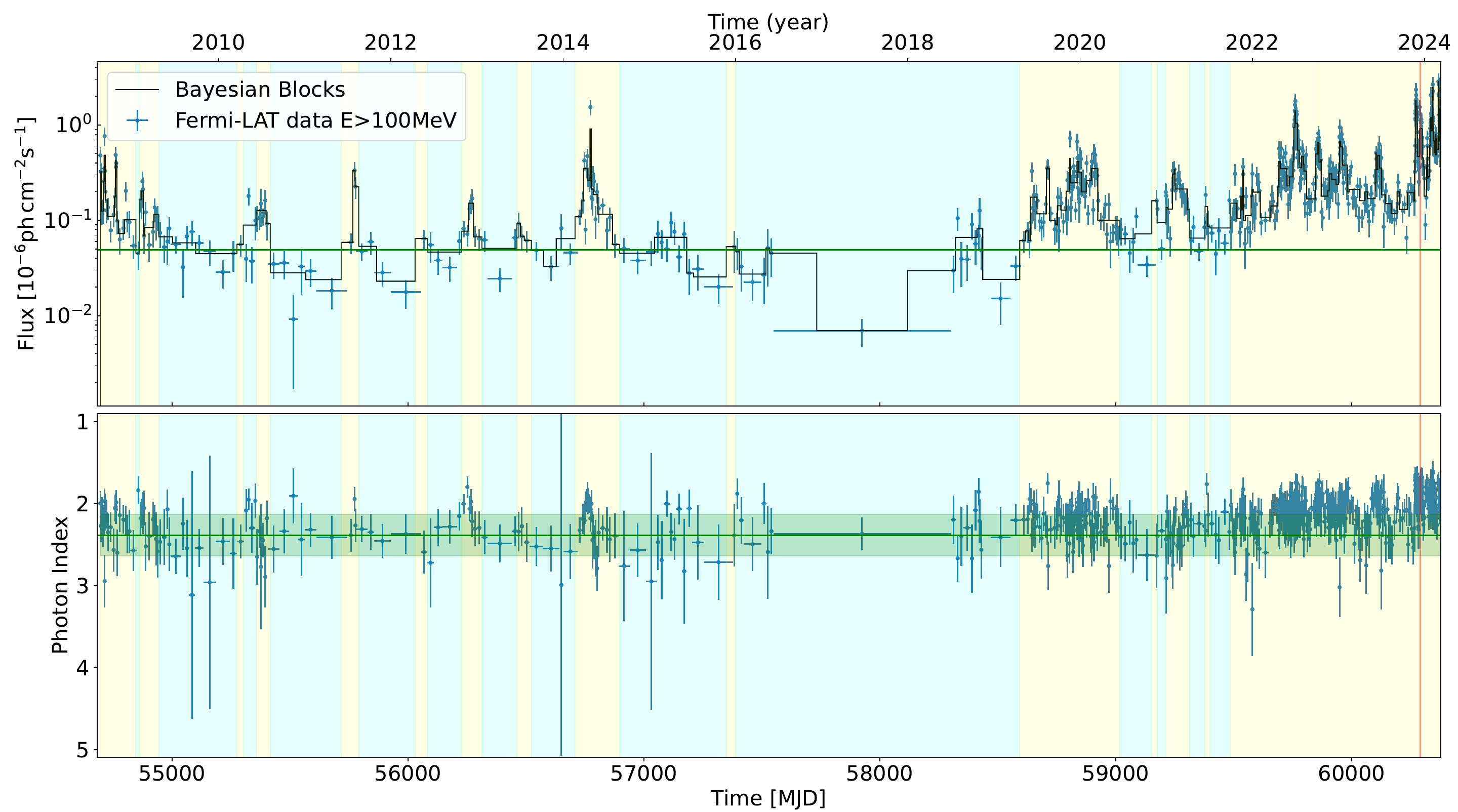}
    \caption{Top: 15-year Fermi-LAT lightcurve of OP 313, divided into quiescent (blue) and flaring (yellow) states. The green line shows the time-weighted average flux of $4.93\times 10^{-8}\, \unit{photons/cm^{2}s}$ computed between the 9th of December 2014 and the 9th of August 2018. Bottom: Fermi-LAT photon index over the same 15-year period. The green line shows the average photon index computed between the 9th of December 2014 and 9th of August 2018 ($\Gamma = 2.39$), with the green band indicating its corresponding $1\sigma$ uncertainty. The red vertical bands highlight the period in which LST-1 detected OP 313.}
    \label{fig:LC-Gamma}
\end{figure*}

By using the adaptive binning algorithm described in \cite{lott_adaptive-binning_2012}, we computed a sequence of time bins spanning from 2008 to 2024 with a fixed relative uncertainty of $0.2$ for the reconstructed \op\, flux.
The lightcurve computed using these adaptive bins is shown in Figure \ref{fig:LC-Gamma}.

Moreover, we identified statistically significant variations of the \op\, flux using the Bayesian Blocks algorithm \citep{scargle_studies_2013}, as implemented in \texttt{astropy}\footnote{\url{https://docs.astropy.org/en/stable/api/astropy.stats.bayesian_blocks.html}}.

The OP 313 flaring activity and quiescent states were determined following the approach in \cite{rodrigues_multiwavelength_2021}. As a proxy for the quiescent state, we considered the time-weighted average flux over the low-activity period spanning from 9 December 2014 to 9 August 2018. The bayesian blocks with fluxes above $4.93\times 10^{-8}\, \unit{photons/cm^{2}s}$ are identified as flaring states, whereas those below this threshold represent the quiescent ones. These are shown in yellow and blue in Figure \ref{fig:LC-Gamma}, respectively. The red band in Figure \ref{fig:LC-Gamma} represents the period in which the LST-1 detected OP 313. The source has remained in a flaring state since late 2021. Figure \ref{fig:LC-Gamma} shows that the photon index from the beginning of 2022 is generally harder than the average measured between the 9th of December 2014 and the 9th of August 2018 ($\Gamma = 2.39 \pm 0.25$, in green). 

\subsection{Neil Gehrels Swift Observatory Data}
\label{Swift}
The \textit{\ac{Swift}} \citep{gehrels_swift_2004} mission is a fast-repositioning, multiwavelength satellite primarily designed for \ac{GRB} studies. It s equipped with three instruments: the Burst Alert Telescope \citep[BAT,][]{barthelmy_burst_2005}, sensitive to the $15-350\,$keV range, the X-Ray Telescope \cite[XRT,][]{burrows_swift_2005}, operating in the $0.2-10\,$keV band, and the UV/Optical Telescope \cite[UVOT,][]{roming_swift_2005}, covering the $190-600\,$nm wavelengths.

We analyzed the \textit{Swift}-XRT observations of \op\, from the 20th of August 2008 to the 9th of March 2024. We used the spectral fitting package XSPEC v12.13.1 \citep{arnaud_xspec_1996} and the python package PyXspec v2.1.4 available in the HEASOFT 

software v.6.33.1, together with the calibration database (CALDB) v20220331. 
We modeled the observed XRT spectra using a \texttt{phabs*zpowerlw} function\footnote{\url{https://heasarc.gsfc.nasa.gov/docs/software/xspec/manual/node128.html}}.
The \texttt{phabs} term accounts for the Galactic extinction, and the hydrogen column density was fixed to $N_{H} = 1.23 \times 10^{20}\, \mathrm{cm^{-2}}$ \footnote{\url{https://heasarc.gsfc.nasa.gov/cgi-bin/Tools/w3nh/w3nh.pl}}. 

Conversely, the \texttt{zpowerlw} model is defined as: 
\begin{equation} \frac{dN}{dE} = N_{0} [E(1+z)]^{\alpha} \end{equation}
where $N_{0}$ is the normalization in units of $\unit{photons/keV/cm^{2}/s}$, $E$ is the energy in keV units, $z$ is the redshift, and $\alpha$ is the photon index.
For each XRT exposure, the intrinsic X-ray SED of \op\, was reconstructed by correcting the observed spectral points for the Galactic extinction effects, using the procedure presented in \cite{abe_very_2025}.

The \textit{Swift}-UVOT data from the same period were processed and analyzed using the \texttt{uvotimsum} and \texttt{uvotsource} tasks in HEASOFT. The available UVOT observations were performed using the optical \textit{v, b, u} and UV \textit{w1, m2, w2} photometric filters. The source counts were extracted from a circular source region of $5$ arcsec radius centered at the \op\, position, whereas a circular nearby region of $20$ arcsec radius devoid of sources was used to derive the background counts. We accounted for the Galactic extinction using $E(B-V)=0.0115\pm0.0005$ \citep{schlafly_measuring_2011}.

\subsection{Optical}

The long-term optical lightcurve was assembled by using the photometric data from several all-sky surveys and dedicated blazar monitoring programs: Zwicky Transient Facility\footnote{\url{https://www.ztf.caltech.edu/}} \cite[ZTF,][]{bellm_zwicky_2018}, Catalina Real-Time Transient Survey\footnote{\url{http://crts.caltech.edu/}} \cite[CRTS,][]{drake_first_2009}, Irsa's Image and Spectrum Server\footnote{\url{https://irsa.ipac.caltech.edu/applications/Atlas/}} (ATLAS), Katzman Automatic Imaging Telescope\footnote{\url{https://www.lickobservatory.org/explore/research-telescopes/katzman-automatic-imaging-telescope/}}\cite[KAIT,][]{filippenko_lick_2001} at the Lick Observatory and the Tuorla blazar monitoring program\footnote{\url{https://users.utu.fi/kani/1m/}} \citep{nilsson_long-term_2018}.

For each of these facilities, the photometric filters included in this work are reported in Table \ref{tab:instruments}.

The survey datasets have been extracted from the corresponding databases, while the dedicated observations of this source from the Tuorla blazar monitoring were analyzed using the standard procedures of differential photometry, with a semi-automatic pipeline \citep[e.g.,][]{nilsson_long-term_2018} and an aperture of $5\,$arcseconds.
The optical data from all the facilities were combined following the prescriptions of \cite{kouch_association_2024}. Data from simultaneous and quasi-simultaneous nights were used to estimate the shifts between different observatories, using the Tuorla blazar monitoring data as reference point.

\subsection{Mets$\ddot{a}$hovi}
We included $37\, \mathrm{GHz}$ observations made with the $13.7\, \mathrm{m}$ diameter Mets\"ahovi radio telescope (see Table \ref{tab:instruments}). The detection limit of the telescope at $37\, \mathrm{GHz}$ is on the order of $0.2\, \mathrm{Jy}$ under optimal conditions. Data points with a signal-to-noise ratio $< 4$ are handled as non-detections. The flux density scale is set by the observations of DR 21, whereas NGC 7027, 3C 274, and 3C 84 are used as secondary calibrators. A detailed description of the data reduction and analysis is given in \cite{terasranta_fifteen_1998}. The error estimate in the flux density includes the contribution from the root mean square (rms) and the uncertainty of the absolute calibration.

\subsection{F-GAMMA, QUIVER and SMAPOL}

The Effelsberg 100-m Radio Telescope measured the total flux density of OP\,313 within the framework of the F-GAMMA (FERMI-GST AGN Multi-frequency Monitoring Alliance) and QUIVER (Monitoring the Stokes \textbf{Q}, \textbf{U}, \textbf{I} and \textbf{V} \textbf{E}mission of AGN jets in \textbf{R}adio) monitoring programs. 
F-GAMMA followed the radio variability of several $\gamma$-ray loud blazars between 2007 and 2015. 
A complete description of the project and the dataset collected by the Effelsberg telescope can be found in \cite{fuhrmann_f-gamma_2016} and \cite{angelakis_f-gamma_2019}. F-GAMMA was followed by the QUIVER monitoring program, which monitors the total flux density and polarization evolution of a subset of the F-GAMMA sources, including OP\,313 on a fortnightly basis since June 2015.

The F-GAMMA and QUIVER observations are performed at several radio bands (depending on receiver availability and weather conditions) from 2.6\,GHz to 44\,GHz (11\,cm to 7\,mm wavelength) using nine receivers located at the secondary focus of the 100-m Radio Telescope. The polarized intensity, position angle, and polarization percentage were derived from the Stokes I, Q, and U cross-scans. The total flux was computed using the calibrators 3C\,286, 3C\,48 and NGC\,7027 (e.g., \cite{myserlis_full-stokes_2018, kraus_intraday_2003}). In this paper, we used the F-GAMMA data from the 22nd of August 2008 to the 1st of June 2014, and the QUIVER data from the 13th of November 2023 to the 30th of March 2024.

The Submillimeter Array (SMA) \citep{ho_submillimeter_2004} was used to obtain total flux at 1.3~mm (225~GHz) within the framework of two complementary programs: the standard SMA flux monitoring campaign \citep{gurwell_monitoring_2007} and the recent SMAPOL (\textbf{S}MA \textbf{M}onitoring of \textbf{A}GNs with \textbf{POL}arization) program. SMAPOL follows the evolution of forty gamma-ray loud blazars, including OP\,313, since June 2022 on a fortnightly cadence. 

We employed  MWC\,349\,A, Callisto, Uranus, Neptune, and Ceres for the total flux calibration according to their visibility.

The SMA data used in this work span from 13 September 2008 to 10 March 2024, and include observations from both the flux monitoring and SMAPOL programs at $225$, $273$ and $350\, \mathrm{GHz}$. 

\subsection{VLBA data}
\label{VLBA}
The VLBA-BU-BLAZAR\footnote{\url{https://www.bu.edu/blazars/}} and \ac{MOJAVE}\footnote{\url{https://www.cv.nrao.edu/MOJAVE/}} monitoring programs consist of monthly \ac{VLBA} observations of blazars and radio galaxies. The MOJAVE data archive consists of nearly ten thousand total density fluxes VLBA observations at $15\unit{GHz}$ ($2\unit{cm}$) of over 500 AGN dating back to 1994 \citep{lister_monitoring_2021}. Instead, the VLBA-BU-BLAZAR monitoring program observes a sample of AGN detected in the $\gamma$-ray band at $43\unit{GHz}$ ($7\unit{mm}$) and $86\unit{GHz}$ ($3.5\unit{mm}$) \citep{jorstad_kinematics_2017, weaver_kinematics_2022}. The available data for OP 313 are taken at $43\unit{GHz}$. 
We included VLBA-BU-BLAZAR data from the 1st of April 2009 to the 22nd of March 2024 and data from the 7th of January 2009 to the 7th of April 2024 from the \ac{MOJAVE} project.

We modeled the sky brightness distributions in the $(u, v)$ visibility plane. We modeled the visibilities of OP 313 with a series of circular Gaussian components using the \texttt{modelfit} task in the Difmap v2.5o software package \citep{shepherd_difmap_1997}. We refer to these Gaussian components using the term “knots”, which are usually compact features of enhanced brightness in the jet. We started with a single circular Gaussian component approximating the brightness distribution of the core, which is typically the brightest feature on the map located close to the base of the jet. Then knots were iteratively added at the locations of bright features. For every knot that we added, we used the \texttt{modelfit} task to determine the best-fit parameters of the knot, according to a $\chi^{2}$ test. The fit procedure was iterated until the addition of a new knot did not significantly improve the $\chi^{2}$ value. The model of a previous epoch was used as the starting model for the next epoch, since we assume that the jet does not change drastically on monthly timescales \citep{weaver_kinematics_2022}. 

\section{Results}
\label{results}
This Section presents the main results of the analyses performed on the multiwavelength dataset.

\subsection{Multiwavelength lightcurves of OP 313}
\label{multi-lightcurve}

\begin{figure*}[htb]
    \centering
    \includegraphics[width=1.0\linewidth]{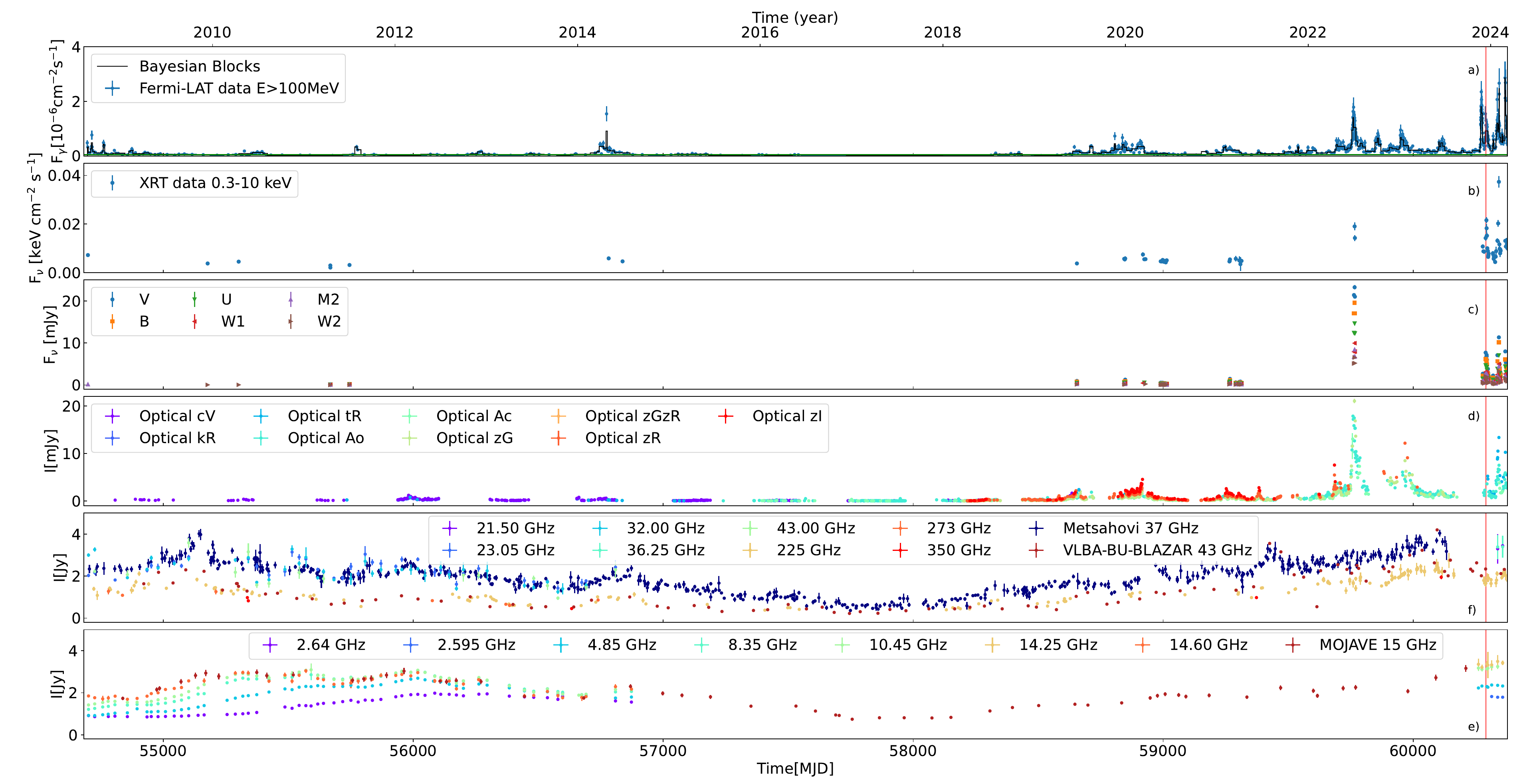}
    \caption{Multiwavelength lightcurve of OP 313. The red vertical band indicates the period when LST-1 detected OP 313. 
    a: \textit{Fermi}-LAT lightcurve at $E>100\,$MeV. 
    b: \textit{Swift}-XRT X-ray lightcurve from 0.3$\, \mathrm{keV}$ to 10$\, \mathrm{keV}$.
    c: \textit{Swift}-UVOT lightcurve in the optical/UV bands.
    d: Optical lightcurve from several facilities and filters: CRTS V-filter (cV), KAIT R-filter (kR), Tuorla R-filter (tR), ATLAS o (Ao) and c-filters (Ac), and Palomar ZTF g, r, and i-filters (zG, zR, and zI). The label zGzR refers to ZTF observations using both g and r filters.
    e: Radio Single Dish \ac{F-GAMMA} lightcurves above 15$\, \mathrm{GHz}$, \ac{SMA} data at 225, 273 and 350$\, \mathrm{GHz}$, including SMAPOL data at 225$\, \mathrm{GHz}$ lightcurves, Mets$\ddot{a}$hovi at 37$\, \mathrm{GHz}$ and VLBA-BU-BLAZAR at 43$\, \mathrm{GHz}$ lightcurves.
    f: Radio Single Dish \ac{F-GAMMA} lightcurves below 15$\, \mathrm{GHz}$ and \ac{MOJAVE} \ac{VLBA} lightcurve.}
    \label{fig:multiwavelengthLC}
\end{figure*}

Figure \ref{fig:multiwavelengthLC} shows the radio to $\gamma$-rays lightcurve from August 4, 2008, to March 9, 2024 (MJD 54682.7 - 60378.0). 

Starting in 2022, the source shows a systematic increase in flux across the $\gamma$-ray, X-ray, and optical/UV bands.
Before this period, its activity was limited to small flaring episodes, marginally visible in the $\gamma$-ray lightcurve. Instead, the radio lightcurves show variability over the 15 years. Across all radio frequencies, a common trend emerges: the source was in a high-flux state from the beginning of our analysis in 2008, followed by a gradual decline until 2019, after which the flux began to increase once more (see more details in Section \ref{Radio_more}). 

\subsection{Selection of the flaring periods of interest}
\label{flares}
\begin{figure*}[htb]
    \centering
    \includegraphics[width=1\linewidth]{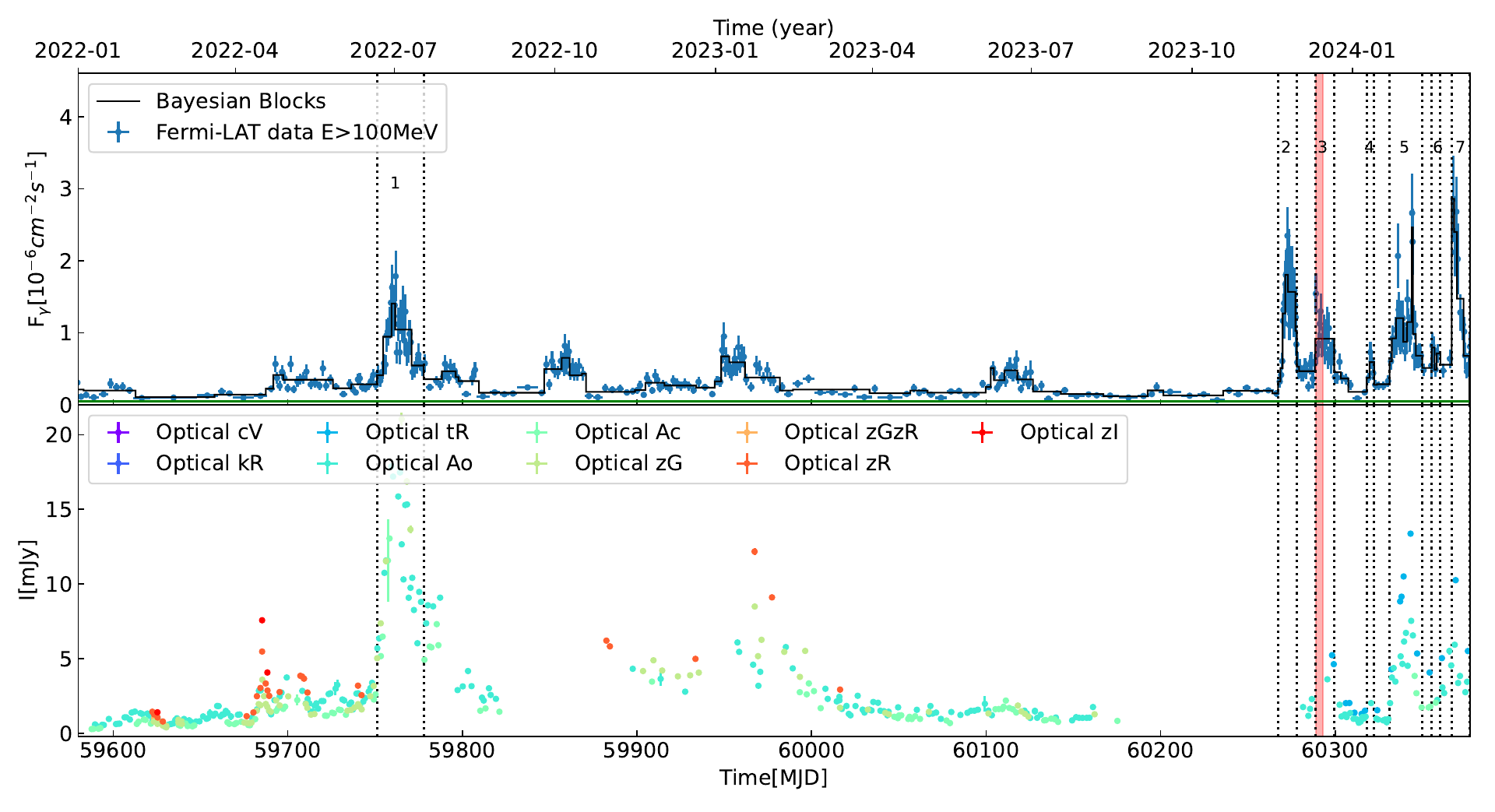}
    \caption{\textit{Fermi}-LAT (top) and Optical (bottom) lightcurves from the 1st of January 2022 to 9th of March 2024. The dashed black lines indicate the brightest $\gamma$-ray flaring periods. The fourth and sixth flaring periods are not between the brightest but are located in time between the ATels \cite{bartolini_atel_2023} and \cite{bartolini_atel_2024}. The red vertical bands indicate the period when LST-1 detected OP 313.}
    \label{fig:LC-zoom}
\end{figure*}

In Figure \ref{fig:LC-Gamma}, the source began a continuous flaring phase at the end of 2021. Particularly noteworthy is the multiwavelength flaring activity, spanning from $\gamma$-rays to optical frequencies, which started in 2022 and persisted until the end of this study. Figure \ref{fig:LC-zoom} presents the $\gamma$-ray and optical lightcurves from the 1st of January 2022 to the 9th of March 2024, showing a good correlation between the two datasets, despite the limited optical coverage during the flaring episode reported in \cite{bartolini_atel_2023}. Further discussion on the $\gamma$-ray–optical correlation is provided in Section \ref{correlation}.

The highest $\gamma$-ray flaring episodes within this time interval were identified by selecting the Bayesian blocks with the highest average flux. Seven distinct flaring episodes were found, marked with black dashed lines in Figure \ref{fig:LC-zoom}. The fourth and sixth flaring periods are not among the brightest, but they occurred temporally between the ATels \cite{bartolini_atel_2023} and \cite{bartolini_atel_2024}. We decided to include them, given our interest in unveiling the mechanisms responsible for the intense flaring activity. Table \ref{tab:flaring_periods} shows: (1) ID of the flaring period, (2) flaring period in MJD, (3) flux peak and its error, (4) photon index corresponding to the flux peak, (5) fluence in the flaring period, defined as the time-integrated flux. We note that although Flare 3 coincides with the detection of OP 313 at energies above 250 GeV, it is not the brightest among the identified events, but it exhibits one of the hardest spectral indices.

\begin{table*}[h]
    \centering
    \caption{Flaring periods of OP 313 starting from January 2022.}
    \begin{tabular}{ccccc}
    \toprule
    ID & Period (MJD) & $F_{peak}\, (\mathrm{ph/cm^{2}/s}) $ & $\alpha_{peak}$ & Fluence ($\mathrm{ph/cm^{2}}$) \\ \midrule
    1 & 59751 - 59778 & $(1.79\pm 0.35)\times 10^{-6}$ & $2.31\pm0.19$ & $1.87\pm0.07$ \\ 
    2 & 60267 - 60278 & $(2.35\pm 0.4)\times 10^{-6}$ & $1.68\pm0.1$ & $1.08\pm0.03$  \\ 
    3 & 60289 - 60299 & $(1.54\pm 0.28)\times 10^{-6}$ & $1.78\pm0.11$ & $0.79\pm 0.04$ \\ 
    4 & 60318 - 60322 & $(7.3\pm 1.41)\times 10^{-7}$ & $1.95\pm 0.14$ & $0.12\pm0.01$ \\ 
    5 & 60331 - 60350 & $(2.65\pm 0.55)\times 10^{-6}$ & $2.15\pm 0.17$ & $1.52\pm0.06$ \\ 
    6 & 60355 - 60360 & $(8.23\pm 1.83)\times 10^{-7}$ & $2.1\pm0.18$ & $0.18\pm 0.02$  \\ 
    7 & 60367 - 60377 & $(2.86\pm 0.6)\times 10^{-6}$ & $1.75\pm0.12$ & $1.22\pm0.1$ \\ \bottomrule
    \end{tabular}
    \tablefoot{The photon index $\alpha_{peak}$ corresponds to the maximum flux $F_{peak}$ in the flaring period.}
    \label{tab:flaring_periods}
\end{table*}

\subsubsection{{Search for} hysteresis patterns}
The hysteresis pattern \citep[e.g][]{kirk_particle_1998} is a loop-like structure that provides insight into the particle acceleration and cooling mechanisms occurring within the blazar's jet. There are two different kinds of hysteresis patterns:
\begin{itemize}
    \item the more common clockwise pattern, indicating the spectral slope is dominated by synchrotron cooling \citep{tashiro_1995pasj47131t_1995};
    \item the rare anti-clockwise pattern, which suggests that the cooling and acceleration rates are comparable and the flaring event propagates from lower to higher energies. 
\end{itemize}

As shown in Figure \ref{fig:LC-Gamma}, the $\gamma$-ray photon statistics obtained with $Fermi$-LAT is large enough to investigate possible hysteresis patterns during the flaring periods defined in Section \ref{flares}. 
We realized the hysteresis plots n which the spectral index is plotted on the y-axis as a function of the flux on the x-axis.

\begin{figure*}[htb]
    \centering
    \subfigure[\label{fig:hysteresis1Flare}]{\includegraphics[width=0.45\linewidth]{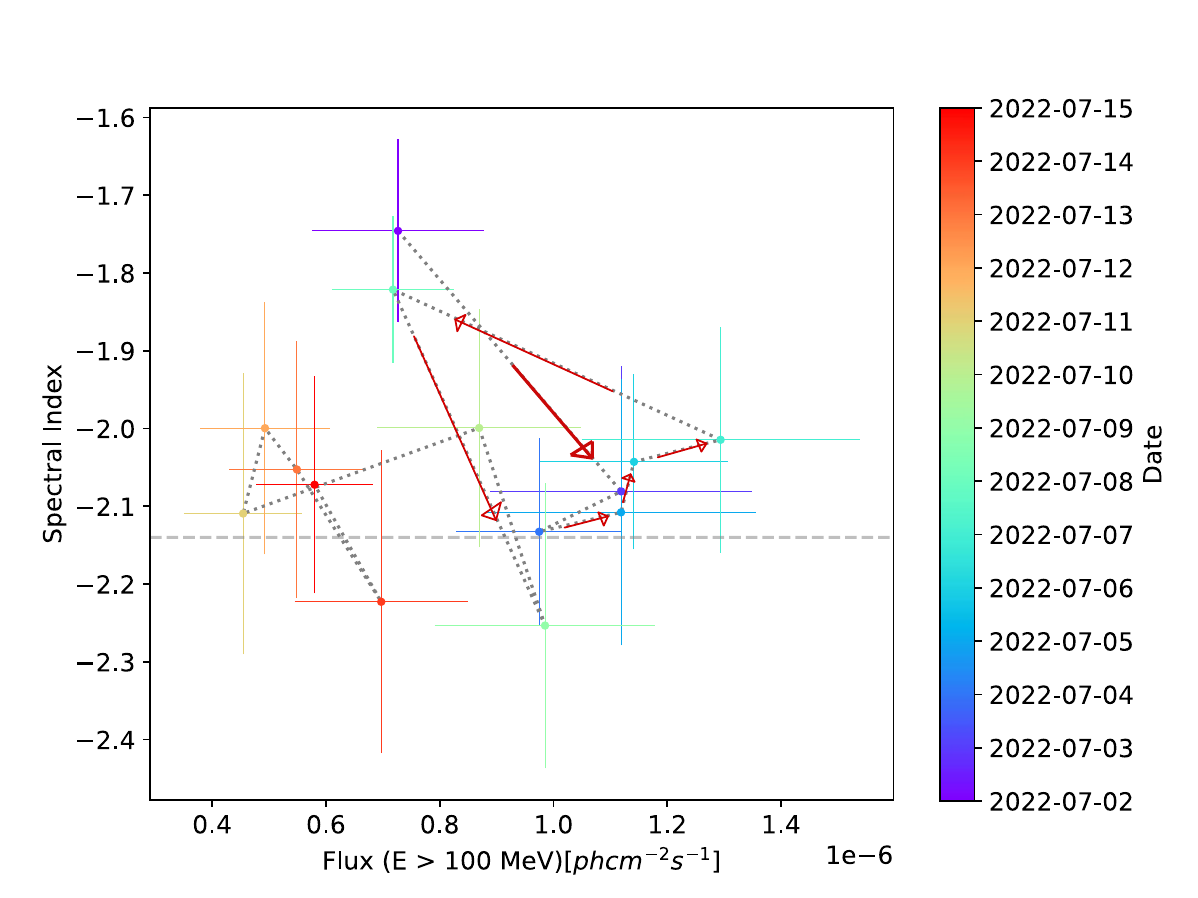}}\quad
    \subfigure[\label{fig:hysteresis5Flare}]{\includegraphics[width=0.45\linewidth]{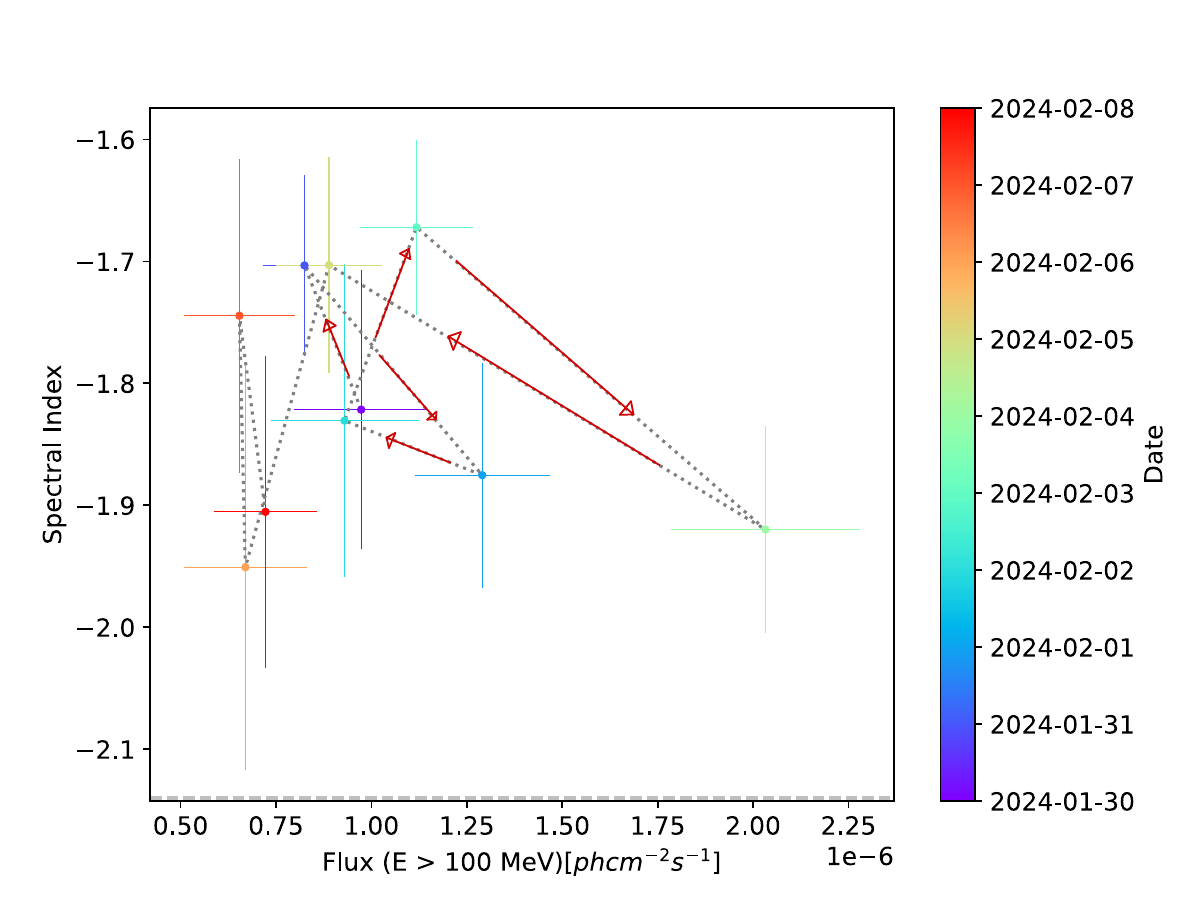}}\quad
    \subfigure[\label{fig:hysteresis2Flare}]
    {\includegraphics[width=0.45\linewidth]{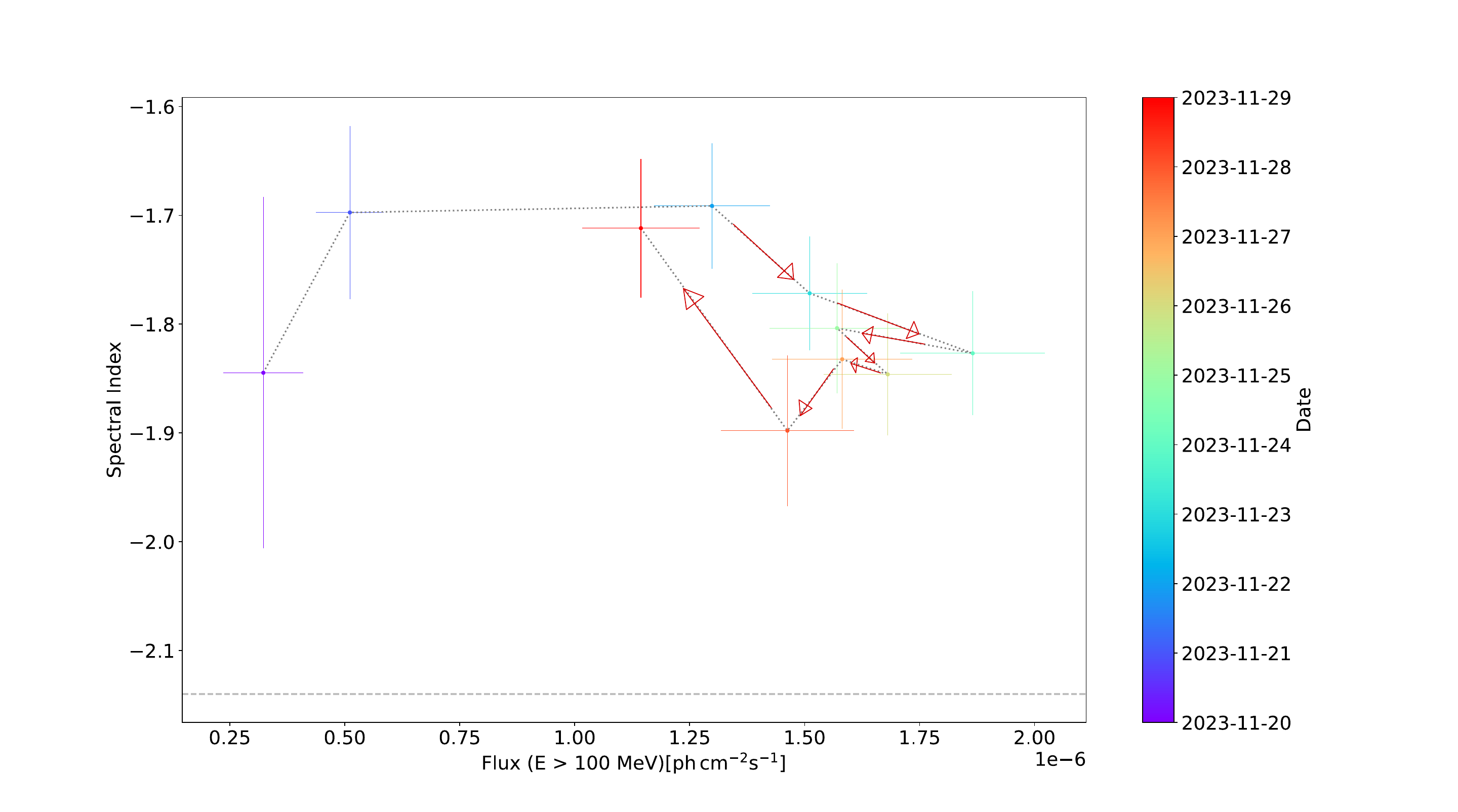}}
    \caption{a) Hysteresis pattern of the first flaring period (2nd of July 2022 - 15th of July 2022). The grey dashed line is included to guide the eye in identifying the temporal sequence of the points. b) Hysteresis pattern of the fifth flaring period (30th of January 2024 - 8th of February 2024). c) Hysteresis pattern of the second flaring period (20th of November 2023 - 29th of November 2023).}    
\end{figure*}

 Visual inspection of the plots (Figure \ref{fig:hysteresis1Flare}) suggests for the 1st flare
 a hint of anti-clockwise pattern 
 arising on the 4th of July 2022. Instead, hints of clockwise hysteresis patterns were found for the second, in Figure \ref{fig:hysteresis2Flare}, and fifth flaring periods, in Figure \ref{fig:hysteresis5Flare}. This means that these flaring periods are dominated by the cooling. 
 There are no hints of hysteresis patterns for the rest of the periods.

Evaluating statistical significance of hysteresis patterns is non-trivial, but one possible test is to bin the data differently.
In Appendix \ref{Appendix hysteresis}, Figure \ref{fig:hysteresis1Flare2days}, \ref{fig:hysteresis5Flare2days} and \ref{fig:hysteresis2Flare2days} present the hysteresis patterns for first, second, and fifth flaring periods, using the average photon index computed in 2-day-long time bins. In this case, the number of data points is smaller as well as the associated uncertainties. The resulting hints of hysteresis pattern are anti-clockwise for the first and fifth flaring periods and clockwise for the second flaring period. Therefore, we conclude that our search for hysteresis patterns was inconclusive even for these extremely bright flares.
 
\subsubsection{Compton dominance}
\label{Compton Dominance} 
Figure \ref{fig:LC-zoom} presents the flaring activity of OP 313 from the 1st of January, 2022, to the 9th of March, 2025. The first optical flaring period is stronger than its $\gamma$-ray counterpart, whereas the opposite trend is observed for the subsequent ones. This disparity between flaring periods across various wavelengths suggests a shift in the origin of the seed photons responsible for Comptonization. For \acp{FSRQ} such as \op, the soft seed photons for external Compton scattering are expected to originate outside the jet \citep{nalewajko_sequence_2017}. Hence, investigating the relative variability between the two \acp{SED} peaks helps to constrain the nature and location of the external photon field. 

\begin{figure*}
    \sidecaption
    \includegraphics[width=12cm]{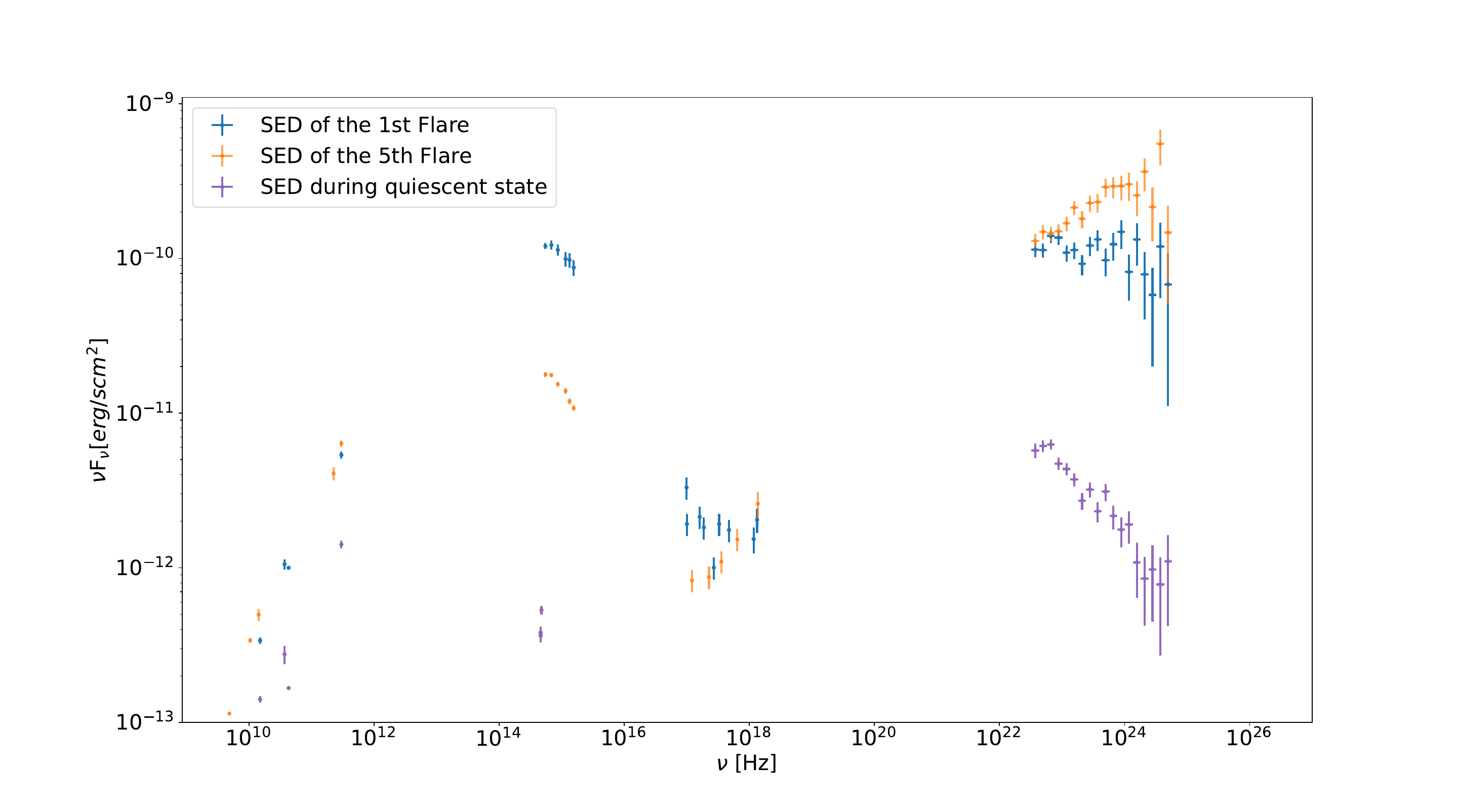}
    \caption{SED of the 1st (2022), in orange, the 5th flaring periods (January 2024), in blue, and a quiescent period that goes from the 1st of March 2016 to the 1st of January 2018, in purple. The X-ray, UV, optical, and radio SED points correspond to the average flux values within each period at their respective frequencies. No \textit{Swift}-XRT and UVOT data were available during the quiescent period.}
    \label{fig:SEDs}
\end{figure*}

In this study, we compared the \acp{SED} of selected flaring periods to investigate differences in the relative intensities of their two peaks. The heights of these peaks in a blazar \ac{SED} are directly connected to the Comptonization processes. When the synchrotron and high-energy peaks have comparable amplitude, the jet is moderately magnetized, and \ac{SSC} dominates the high-energy emission \citep{finke_compton_2013, potter_synchrotron_2013}. When the high-energy peak significantly exceeds the synchrotron peak, the Compton dominance is high, and \ac{EC} scattering becomes the primary radiative mechanism. A high synchrotron luminosity ($L_{syn} > 10^{47} \unit{erg/s}$, \cite{janiak_magnetization_2015}) implies highly magnetized emitting region in the jet and efficient synchrotron cooling \citep{janiak_magnetization_2015}. Figure \ref{fig:SEDs} shows the \acp{SED} of the first and fifth flaring periods, as well as a quiescent period going from the 1st of March 2016 to the 1st of January 2018. The first flaring period \ac{SED} exhibits peaks of comparable heights, whereas the fifth presents peaks at different heights. This indicates low Compton dominance during the first flaring period and a more Compton-dominated scenario in the fifth. All the flaring periods from November 2023 to March 2024 exhibit enhanced Compton dominance. A lower Compton dominance suggests that \ac{SSC} emission is dominating over the \ac{EC}, whereas a higher dominance is often interpreted as evidence for prevailing \ac{EC} scattering \citep{sikora_comptonization_1994}. In Section \ref{SED}, for coherence, we discuss in detail the \ac{SED} modeling study of these 2 flaring periods.

\subsubsection{Multi-band correlation}
\label{correlation}
Understanding blazar jet emission and its dynamics requires examining how changes in their physical properties drive variability across various frequencies \citep{mohanaa_multiband_2024}. A cross-correlation approach effectively identifies plausible links across different blazar emission bands, distinguishes emission components in the jet, constrains the emission zone's location, and differentiates between \ac{SSC} and \ac{EC} models \citep{liodakis_probing_2019}. 

\begin{figure*}[ht]
    \centering
    \subfigure[]{\includegraphics[scale=0.45]{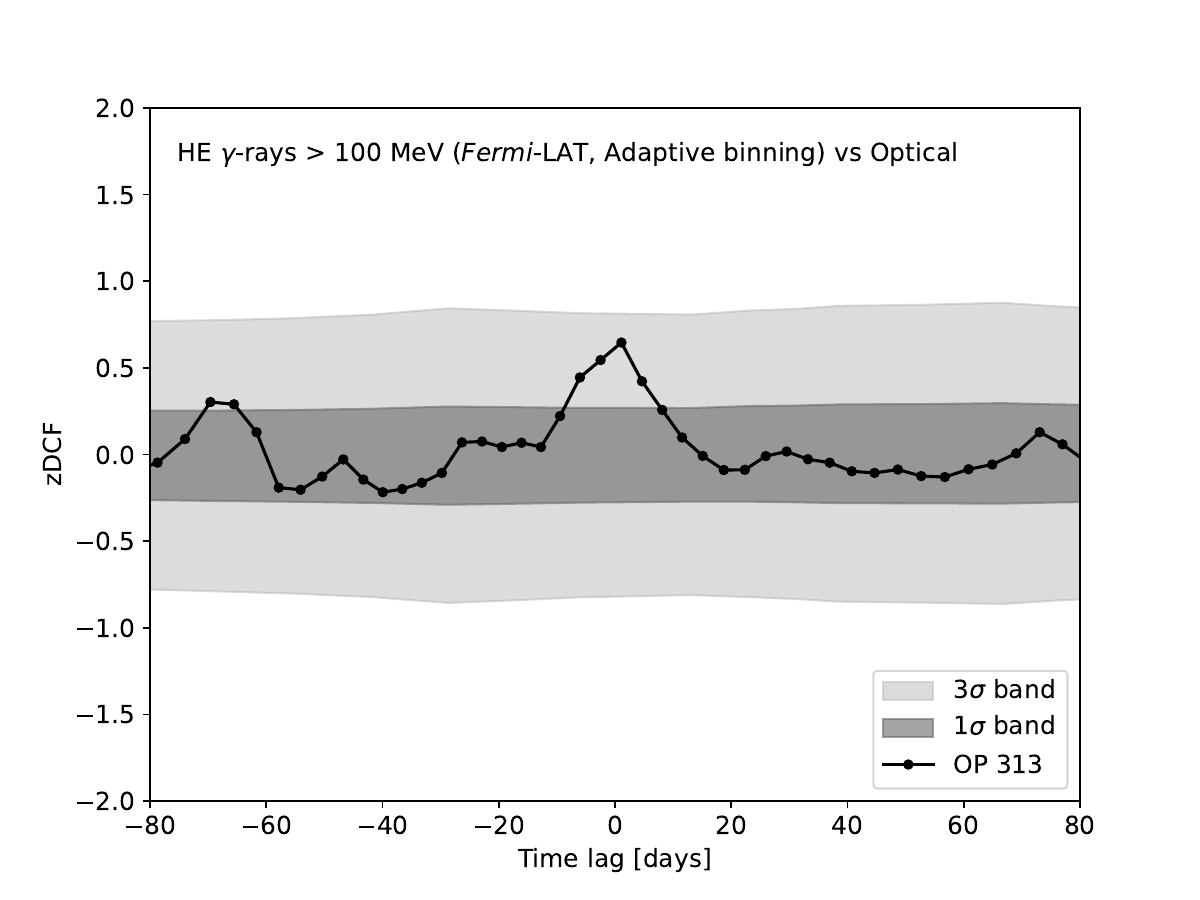}\label{fig:Correlation}}
    \subfigure[]{\includegraphics[scale=0.45]{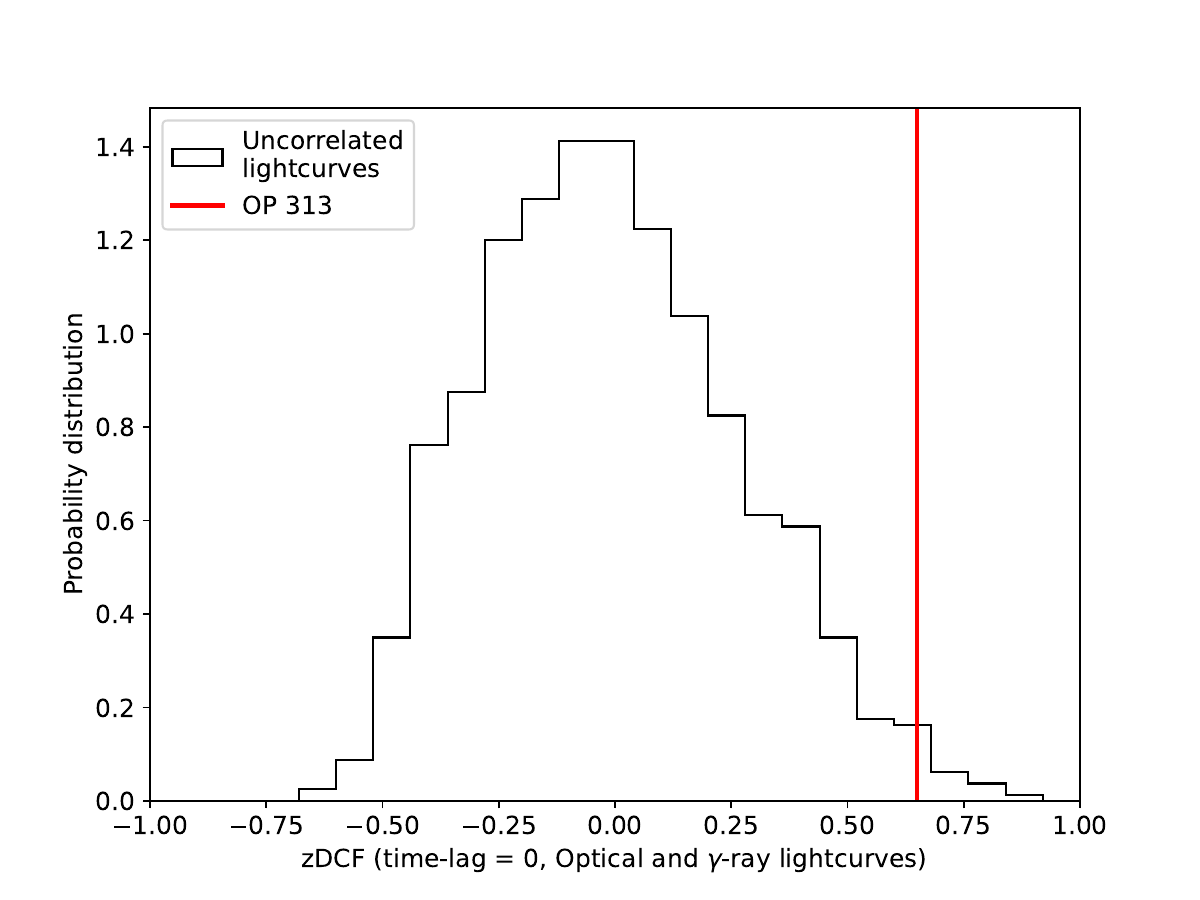}\label{fig:Correlation-histogram}}
    \caption{(a) zDCF analysis on the short-term period of OP 313 optical and $\gamma$-ray lightcurves; (b) distribution of the reconstructed zDCF at zero time-lag for $10^{4}$ simulated uncorrelated $\gamma$-ray and optical lightcurves. The red line marks the reconstructed zDCF value with the observed OP 313 lightcurves in Figure \ref{fig:LC-zoom}.}
    \label{fig:Crosscorrelation}
\end{figure*}

The z-transformed Discrete Correlation Function \citep[zDCF][]{alexander_is_1997} was employed to estimate the cross-correlation between the $\gamma$-ray and optical lightcurves \citep[e.g., as in][]{abe_very_2025}. The zDCF evaluates the delay-dependent multi-band cross-correlations. This method is more sophisticated than the Discrete Correlation Function \citep[DCF;][]{edelson_discrete_1988} because it uses equal-population binning and Fisher’s z-transform, and it is developed for sparse and unevenly sampled astronomical time series. A Python implementation of the original Fortran zDCF algorithm called the pyZDCF package \citep{jankov_pyzdcf_2022} was used to evaluate the zDCF. We investigated the cross-correlation between the optical and $\gamma$-ray datasets both in the long-term period (see Figure \ref{fig:multiwavelengthLC}) and in the short-term period from 2022-2024 (see Figure \ref{fig:LC-zoom}). The zDCF evaluation was performed using bins of delay of 2, 3, and 4 days. For all the delay bin sizes we used, the zDCF showed a peak at a delay compatible with 0 days. For the long-term (short-term) period, the zDCF at zero delay was found to be equal to $0.8$ ($0.6$, see Figure \ref{fig:Correlation}).

To perform the significance analysis of the observed lightcurves' zDCF, we modeled the power spectral density (PSD) of each observed lightcurve using a power-law function $PSD(\omega) = A\omega^{-\beta}$. We estimated a best-fit index value for the $\gamma$-ray lightcurve of $\beta_{\gamma} = 1.2$, in agreement with \cite{tarnopolski_comprehensive_2020}, whereas for the optical R-band lightcurve, we found $\beta_{\textit{OPT}} = 1.2$, which is comparable with the typical ranges reported in \cite{nilsson_long-term_2018} for blazars.

The significance of the observed zDCF peak was evaluated by generating $10^{4}$ pairs of uncorrelated $\gamma$-ray and optical lightcurves with the same PSDs and flux distributions as the real ones. We used the DELCgen Python package \citep{connolly_asclnet_2016} to generate the fake lightcurves population using the \cite{emmanoulopoulos_generating_2013} (EM) method. We used the EM method because it also takes into account the distribution of the observed lightcurves' fluxes, whereas the \cite{timmer_generating_1995} (T$\&$K) method simulates lightcurves assuming the observed fluxes follow a Gaussian distribution.

These simulated pairs of lightcurves were used to derive the confidence bands of the zDCF under the hypothesis of non-correlation. The $1\sigma$ and $3\sigma$ bands for the 2022-2024 period are shown in Figure \ref{fig:Correlation}. In the bin of zero delay, we found the positive correlation with zero delay between the optical and $\gamma$-ray lightcurves to be significant above the 95\% confidence level, corresponding to a confidence level of roughly $2\sigma$ (see Figure \ref{fig:Correlation-histogram}).
 
This correlation suggests a common origin for the flaring emission in the two bands, indicating that leptonic processes primarily dominate the $\gamma$-ray emission, though a subdominant hadronic contribution cannot be ruled out \citep{acharyya_investigating_2023}. 
 
\subsection{\ac{VLBA} jet kinematics results}
The \ac{VLBA} analysis consisted of the analysis of data from the \ac{MOJAVE} and VLBA-BU-BLAZAR programs. We identified new knots in the parsec-scale jet that may be responsible for the flaring activity observed in OP 313. To interpret these multiwavelength flaring events, we performed a kinematic analysis of the newly detected components.

\subsubsection{MOJAVE VLBA analysis results}
\label{Mojave}

\begin{figure*}
    \sidecaption
    \includegraphics[width=12cm]{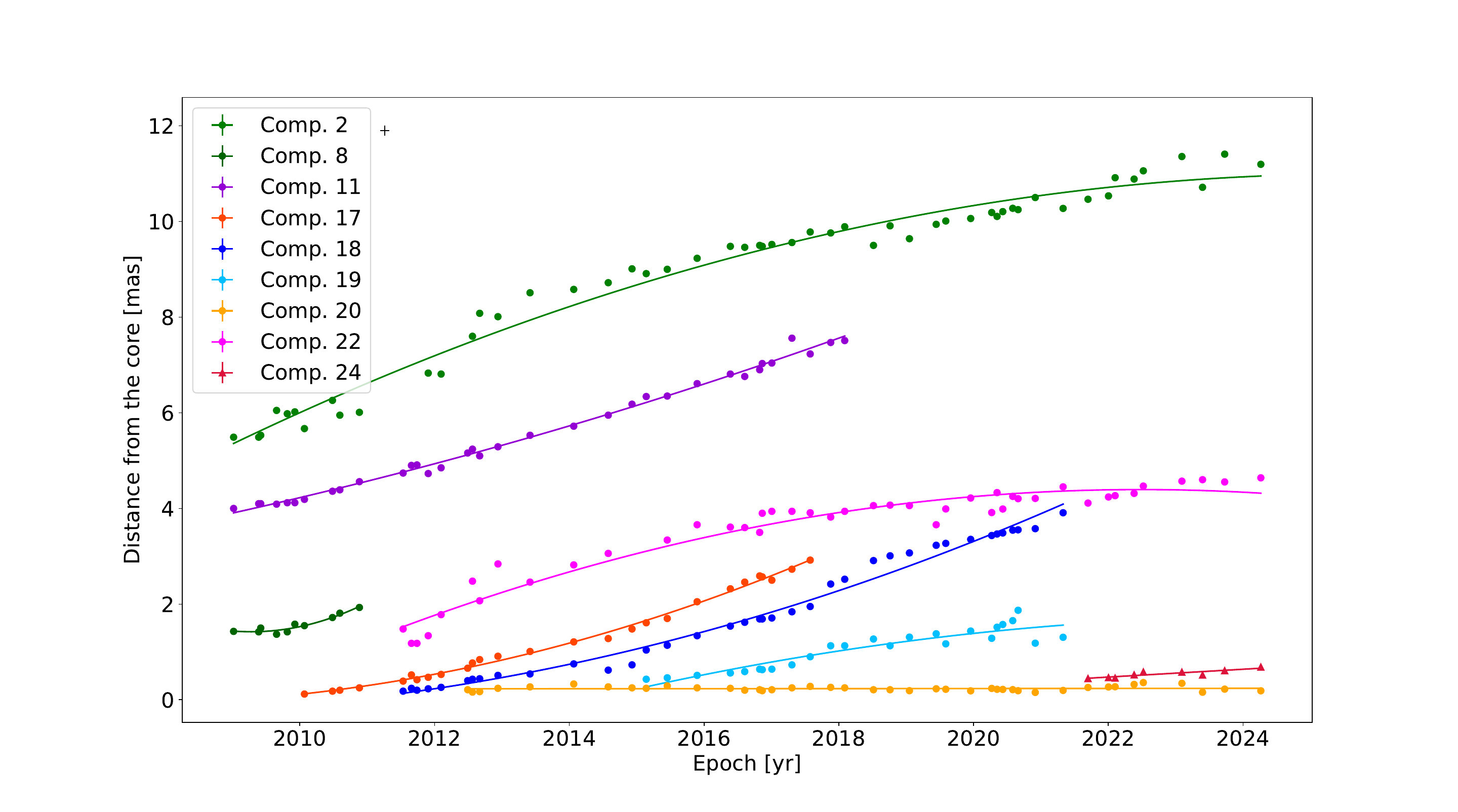}
    \caption{Angular separation from the core versus time for Gaussian jet features of the \ac{MOJAVE} data. Colored symbols indicate robust features for which kinematic fits were obtained. The $1\sigma$ positional errors on the individual points typically range from $10\%$ of the FWHM restoring beam dimension for isolated compact features to $20\%$, represented as a black cross in the plot, of the FWHM for weak features \citep[see e.g.,][]{lister_mojave_2009, lister_mojave_2019, lister_monitoring_2021}.}
    \label{fig:MOJAVE-components}
\end{figure*}

The jet components and corresponding kinematic parameters of OP 313 between 1995 and August 2019 have been previously presented in \cite{lister_mojave_2019} and \cite{lister_monitoring_2021}. Since our long-term multiwavelength study extends to March 2024, we analyzed all the public visibilities up to the 4th of April 2024, which is the closest observation to the 9th of March, employing the procedure outlined in Section \ref{VLBA}. The main outcomes of this analysis are:
\begin{itemize}
    \item The visibility modeling results obtained between August 2008 and August 2019 are in agreement with those reported in \cite{lister_mojave_2019} and \cite{lister_monitoring_2021}. Hence, in Figure \ref{fig:MOJAVE-components}, the components and their distance from the core in this period are the same as reported in the literature;
    \item A new, well-defined jet component is first detected on 13 April 2021.
\end{itemize}

We employed the same approach as \cite{lister_monitoring_2021} to determine if this new component is a robust cross-identification, checking whether the feature was present across at least 5 epochs, and considering the consistency of its sky position and flux density over time. The new component was named ``Component 24'' and is colored in red in Figure \ref{fig:MOJAVE-components}. 

We analyzed the kinematics of all individual robust jet features, from August 2008 to April 2024, employing the method reported in \cite{lister_monitoring_2021}. We fitted the sky positions of these features, as follows:
\begin{equation}
\label{eq:3}
    x(t) = x_{mid} + \mu_{x}(t-t_{mid}) + \frac{\dot{\mu}_{x}}{2}(t-t_{mid})^{2}
\end{equation}
\begin{equation}
\label{eq:4}
    y(t) = y_{mid} + \mu_{y}(t-t_{mid}) + \frac{\dot{\mu}_{y}}{2}(t-t_{mid})^{2} 
\end{equation}
where $t_{mid}$ represents the midpoint between the first and last observation epochs of the component, $\mu_{x}$ and $\mu_{y}$ are the fitted angular velocities in each sky direction, and $\dot{\mu_{x}}$ and $\dot{\mu_{y}}$ are the fitted accelerations. The $1\sigma$ uncertainty on the position of the knots ranges from $10\%$ of the full-width half maximum (FWHM) restoring beam dimension (a median reference value is 0.096 along the major axis of the restoring beam and 0.056 along its minor axis) for isolated compact features to $20\%$ of the FWHM for weak features (around 0.19 along the major axis and 0.11 along the minor axis) \citep[see e.g.,][]{lister_mojave_2009, lister_mojave_2019, lister_monitoring_2021}. The acceleration of the new ``Component 24'' could not be evaluated because it does not meet the requirement of being detected in at least ten epochs. Instead, ``Component 20'' corresponds to a static shock. To determine the proper motion vector modulus and apparent velocity in units of speed of light, we used the following equations:
\begin{equation}
\label{mu}
\mu = \sqrt{\mu_{x}^{2} + \mu_{y}^{2}}
\end{equation}
\begin{equation}
\label{beta}
    \beta_{app} = \mu D_{A} (1+z)
\end{equation}
where $D_{A}$ corresponds to the angular distance and z is the redshift of the source.

Table \ref{tab:MOJAVE} in Appendix \ref{Appendix radio} presents the parameters derived from the fitting procedure, which characterize the kinematics of the jet components.

\subsubsection{VLBA-BU-BLAZAR analysis results}
\begin{figure*}
    \sidecaption
    \includegraphics[width=12cm]{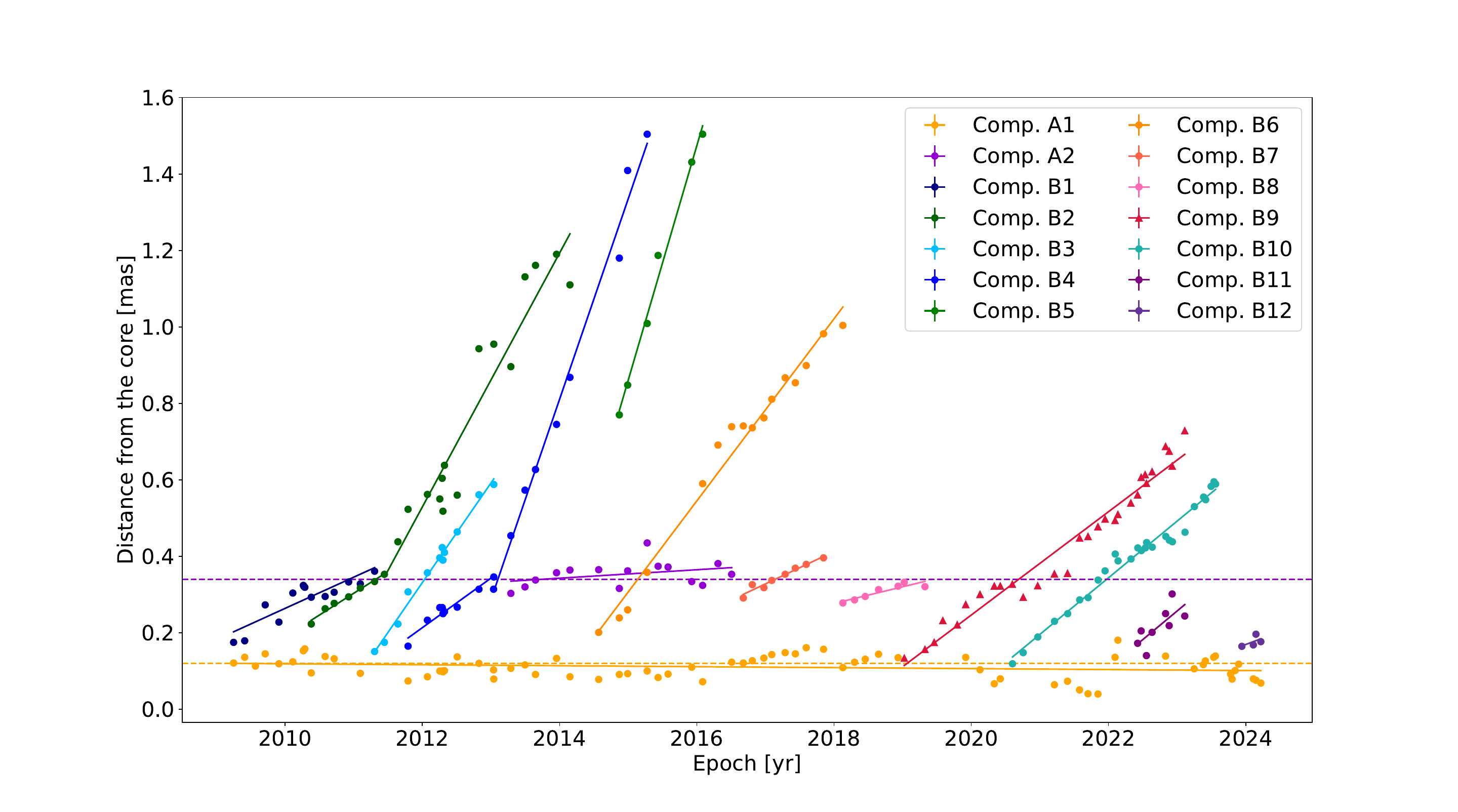}
    \caption{Angular separation from the core versus time for Gaussian jet features of the \ac{VLBA-BU-BLAZAR} data. Colored symbols indicate robust features for which kinematic fits were obtained. The 2 dashed lines in yellow and purple represent the static shocks' mean distance from the core.}
    \label{fig:Boston_plot}
\end{figure*}

The VLBA-BU-BLAZAR jet components and corresponding kinematic parameters of OP 313 between 2009 and 2018 have been presented in \cite{jorstad_kinematics_2017} and \cite{weaver_kinematics_2022}. We analyzed the public visibilities up to the 7th of April 2024 employing the method reported in Section \ref{VLBA}. The main outcomes
of this analysis are:
\begin{itemize}
    \item The visibility modeling results obtained for the period between 2008 and 2018 are consistent with those presented in  \cite{jorstad_kinematics_2017} and \cite{weaver_kinematics_2022}. Hence, Figure \ref{fig:Boston_plot} reproduces the components ``A1'', ``A2'' and from ``B1'' to ``B8'', along with their distances from the core as reported in the literature;
    \item A new set of jet components arose starting from 2019, as shown in Figure \ref{fig:Boston_plot}. 
\end{itemize}
The components' flux and coordinates uncertainties were computed using the formalism described in \cite{jorstad_kinematics_2017} and are presented in Table \ref{tab:Boston}. This formalism is based on an empirical relation between the uncertainties and the brightness temperatures of knots: $T_{b,obs} = 7.5 \times 10^{8}S/a^{2}\,\mathrm{K}$, where $T_{b,obs}$ is the brightness temperature of the knot in Kelvin, $S$ is the flux density in $\mathrm{Jy}$ and $a$ is the angular size in $\mathrm{mas}$ of the knot \citep[e.g.,][]{jorstad_polarimetric_2005, casadio_multi-wavelength_2015, weaver_kinematics_2022}. The uncertainties are calculated as follows: the $1\sigma$ uncertainty on the x-axis is $\sigma_{x}\approx 1.3 \times 10^{4}\, T_{b,obs}^{-0.6}$, the uncertainty on the y-axis is $\sigma_{y}\approx 2\sigma_{x}$, the flux uncertainty is $\sigma_{S}\approx 0.09\,T_{b,obs}^{-0.1}$ and the angular size uncertainty is $\sigma_{a} = 6.5 \times T_{b,obs}^{-0.25}$. We added a minimum positional error of $0.005\, \mathrm{mas}$ (related to the resolution of the observations) and a typical amplitude calibration error of $5\%$ \citep{jorstad_kinematics_2017, weaver_kinematics_2022} to $\sigma_{x}$ and $\sigma_{y}$.

The kinematic properties of the jet components were derived following \cite{weaver_kinematics_2022}. Components detected in more than four epochs, exhibiting consistent positions and fluxes over time, were fitted using the weighted least-squares implementation in the \texttt{statsmodels} package \footnote{\url{https://www.statsmodels.org/stable/index.html}} \citep{seabold_statsmodels_2010}. From the comparison between the $\chi^{2}$ obtained in the fitting process and $\chi^{2}_{crit}$ at a significance level of $\zeta=0.05$ values, we identified accelerating components ($\chi^{2}$>$\chi^{2}_{crit}$), which required segmented fits representing distinct velocities along their motion. Accordingly, Table \ref{tab:Boston}, in Appendix \ref{Appendix radio}, lists some components with two angular proper motion and apparent speed entries for components fitted with two segments. The derived jet structure and kinematic parameters are consistent with \cite{jorstad_kinematics_2017} and \cite{weaver_kinematics_2022}. The parallel and perpendicular accelerations are given by: $\dot{\mu}_{\parallel}=\dot{\mu}_{x}\sin\langle \Theta_{jet}\rangle + \dot{\mu}_{y}\cos\langle \Theta_{jet}\rangle$ and $\dot{\mu}_{\perp}=\dot{\mu}_{x}\cos\langle \Theta_{jet}\rangle - \dot{\mu}_{y}\sin\langle \Theta_{jet}\rangle$, where $\langle \Theta_{jet}\rangle =$ $-64.9 \pm 4.7$ is the average jet position angle for \op\, \citep{weaver_kinematics_2022}, describing the jet orientation on the sky plane, measured from north to east \citep{lister_mojave_2009}. Table \ref{tab:Boston} reports the kinematic properties of the new components arising after 2018. Notably, ``Component 24'' appears consistent with the component ``B9'', as both occupy similar distances from the core and exhibit comparable proper motion magnitudes and apparent speeds, both in red in Figures \ref{fig:MOJAVE-components} and \ref{fig:Boston_plot}. Further observations are required to robustly confirm the identification of some of the later components, such as ``B12'', which appears in only four epochs in our dataset, as well as others that emerged toward the end of our analysis. For more comprehensive details and information about later components, we refer the reader to Jorstad et al.'s forthcoming publication.

\subsubsection{Lorentz Factor and Viewing Angle of Jet Components}
\label{factorsandangles}
Blazar jets are known for their extraordinarily fast variability, boosted emission, and apparent superluminal motion of jet components, due to relativistic processes that dominate the jet's emission. These relativistic effects are quantified by Lorentz factor ($\Gamma$), the Doppler factor ($\delta$), and viewing angle ($\Theta_{0}$), which is the angle between the jet axis and the line of sight to the observer \citep{hovatta_doppler_2009}. The apparent velocities of the new components were used to compute their corresponding Lorentz factors and viewing angles following the formalism of \cite{jorstad_kinematics_2017} and \cite{weaver_kinematics_2022}: 
\begin{equation}
    \label{gamma}
    \Gamma = \frac{\beta_{app}^{2} + \delta_{var}^{2} + 1}{2\delta_{var}}
\end{equation}
\begin{equation}
    \label{theta}
    \tan \Theta_{0} = \frac{2\beta_{app}}{\beta_{app}^{2} + \delta_{var}^{2} - 1}
\end{equation}

where $\delta_{var}$ is the variability Doppler factor and is defined as:
\begin{equation}
    \delta_{var} \approx \frac{16 s D_{\unit{Gpc}}}{\tau_{var}(1+z)}
\end{equation}
where $D_{\unit{Gpc}}$ is the luminosity distance of OP 313 in $\unit{Gpc}$, $s = 1.6a$ is the diameter of a face-on disk with $a$ being the angular size of the knot in $\mathrm{mas}$, which is the FWHM of a Gaussian brightness distribution, and $\tau_{var} = |1/k|$ is the variability timescale. The parameter $k$ is obtained by fitting the flux decay of the knots with an exponential model $\ln(S(t)/S_{0}) = k(t-t_{max})$, where S is the flux density, $S_{0}$ is the flux fitted at  $t_{max}$, and $k$ is the slope \citep{weaver_kinematics_2022}. Using this formalism, we computed the Doppler factor, Lorentz factor, and viewing angle for the new components, from ``B9'' to ``B12''; the results are listed in Table \ref{tab:viewing_angle}. For the other components, these parameters are available in \cite{weaver_kinematics_2022}. The Doppler factors of the knots from ``B9'' to ``B12'' range between 8 and 15, while the Loretz factors range between 7 and 13. These values are in agreement with the general trend for \acp{FSRQ} reported in \cite{hovatta_connection_2014} and \cite{liodakis_constraining_2018}. Using the equation \ref{theta}, the inferred viewing angles are between 2 and 7 $\, \mathrm{deg}$.The accuracy for the component ``B12'' is limiteddue to its detection in only a few epochs. Future analysis of additional VLBA-BU-BLAZAR visibilities, as well as the upcoming paper from Jorstad et al. {\it in prep.}, may improve these estimates.

\begin{table*}
\centering
\caption{Doppler factor ($\delta_{var}$), Lorentz factor ($\Gamma$), and viewing angle ($\Theta_{0}$) of the new components detected from 2019 in VLBA-BU-BLAZAR visibilities.}
    \begin{tabular}{ccccc}
    \toprule
         ID  & B9 & B10 & B11 & B12 \\
    \midrule
    $\delta_{var}$ & $12.94 \pm 2.77$ & $14.95 \pm 0.68$ & $7.8 \pm 0.99$ & $24.86 \pm 1.55$ \\
    $\Gamma$ & $8.12 \pm 0.37$ & $8.05 \pm 0.54$ & $6.83 \pm 0.36$ & $12.51 \pm 0.50$ \\
    $\Theta_{0}$ (\unit{deg}) & $3.55 \pm 1.22$ & $1.93 \pm 0.39$ & $7.29 \pm 1.07$ & $0.32 \pm 0.05$ \\
    \bottomrule
    \end{tabular}
\tablefoot{The values are computed following \cite{weaver_kinematics_2022}.}
\label{tab:viewing_angle}
\end{table*}

Moreover, using the Doppler factor at 37$\, \mathrm{GHz}$ $\delta_{var} = 13.9^{+7.9}_{-3.7}$\citep{liodakis_identifying_2021}, we recalculated the viewing angle and  Lorentz factor to be compared with the values obtained in (see Table \ref{tab:viewing_angle_1}). The results are broadly consistent with Table \ref{tab:viewing_angle}, except for ``B12'', which has limited data.

\begin{table*}
\centering
\caption{Lorentz factor ($\Gamma_{var}$) and viewing angle ($\Theta_{0}$) of the new components arisen from 2019 in the visibilities of the VLBA-BU-BLAZAR project.}
    \begin{tabular}{ccccc}
    \toprule
         ID  & B9 & B10 & B11 & B12 \\
    \midrule
    $\Gamma_{var}$ & $8.47^{+0.17}_{-0.72}$ & $7.55^{+0.41}_{-0.63}$ & $8.58^{+0.1}_{-0.73}$ &  $7.08^{+0.46}_{-0.52}$ \\
    $\Theta_{0}$ (\unit{deg}) & $3.17^{+2.96}_{-1.4}$ & $2.22^{+2.37}_{-1.17}$ & $3.24^{+2.85}_{-1.35}$ & $1.02^{+1.15}_{-0.55}$ \\
    \bottomrule
    \end{tabular}
\tablefoot{The values reported here are obtained using the Doppler factor reported in \cite{liodakis_identifying_2021} $\delta_{var} = 13.87^{+7.87}_{-3.72}$.}
\label{tab:viewing_angle_1}
\end{table*}

Overall, the inferred values in Table \ref{tab:viewing_angle} and in Table \ref{tab:viewing_angle_1}  for B9–B12 are consistent with the general trends for \acp{FSRQ}, where $\delta_{var} < 40$ and $\Gamma_{var}\sim 10$ for \ac{FSRQ} \citep{hovatta_connection_2014, liodakis_constraining_2018}. 

\subsection{Search for the origin of the $\gamma$-ray flaring periods}
\label{Radio_more}
In Figure \ref{fig:multiwavelengthLC}, the radio flux was increasing until 2010 and then started rising again around 2019. Although the lightcurve suggests a possible long-term trend, previous studies of the radio variability of \op\ using Mets\"ahovi data from 1985 to 2023 show that no characteristic variability timescale can be well-established for this source \citep[][]{kankkunen_long-term_2025}. Consequently, the flaring activity observed since 2022 cannot be associated with any well-defined periodic behavior.

\begin{figure*}
    \centering
    \includegraphics[width=1.\linewidth]{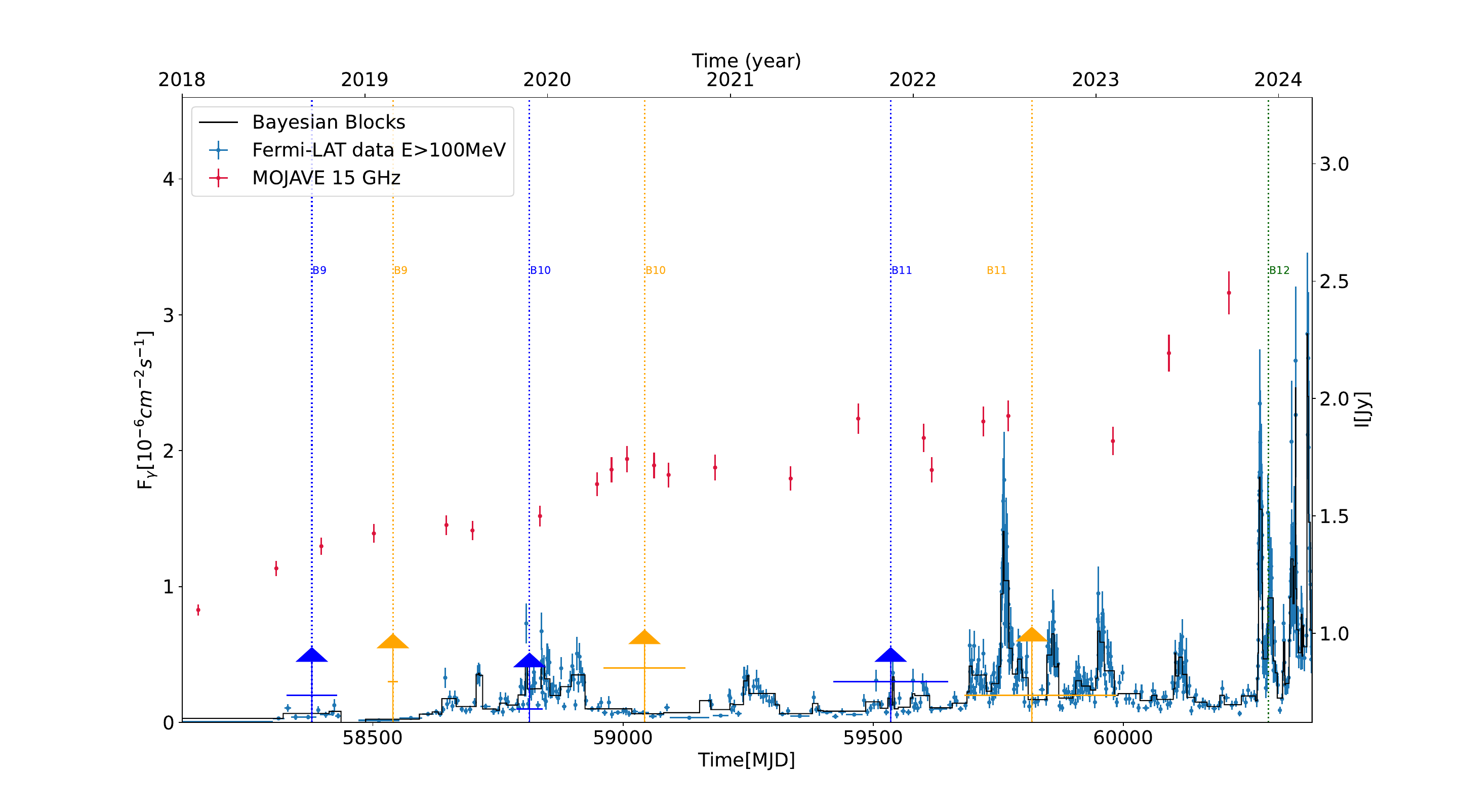}
    \caption{Comparison between the Fermi-LAT data (in blue) with their Bayesian blocks (in black) and the total flux density of \ac{MOJAVE} data at $15\, \mathrm{GHz}$ (in red). The dashed blue lines represent the time of ejections for different knots, found in the \ac{VLBA-BU-BLAZAR} visibilities, and the blue lines are the corresponding errors, while the orange dashed lines represent the epochs when the knots encounter the standing shock ``A1'' with the corresponding errors. In green, the epoch in which ``B12'' was found.}
    \label{fig:0_separation}
\end{figure*}

Figure \ref{fig:0_separation} compares the \ac{Fermi-LAT} and \ac{MOJAVE} total flux density lightcurves from 2018 to March 2024, suggesting a time delay between the two bands. This time lag was investigated for OP 313 in \cite{kramarenko_decade_2022}, who reported a delay of $96^{+16}_{-11}$ days between $\gamma$-ray and radio. A positive time lag indicates that the $\gamma$-ray emission leads the radio emission, implying that the region responsible for the $\gamma$-ray flares is closer to the central engine than the radio-emitting region. The \cite{kramarenko_decade_2022} time lag is smaller than the value we derived looking at the $\gamma$-ray and radio emissions between 2020 and 2021, shown in Figure \ref{fig:0_separation}, where we measured a delay of $201\pm 16$ days by comparing the difference between the epochs where the radio and $\gamma$-ray emissions had a peak. Furthermore, \cite{kramarenko_decade_2022} computed the distance between the $\gamma$-ray emission region and the central engine, taking the radio core opacity as the main source of the time lag. For \op, they found that the $\gamma$-ray emission region is located at $4.96^{+12.53}_{-12.96}\unit{pc}$ \citep{kramarenko_decade_2022} from the central engine, beyond the \ac{BLR} whose external radius is $\simeq 0.1\, \mathrm{pc}$ (see Section \ref{SED} for the derivation of the \ac{BLR} radius). This motivated us to foresee a photon field responsible for the \ac{EC} in the dusty torus, outside the \ac{BLR}. 

\begin{table*}
    \centering
    \caption{Epoch of ejection $T_{0}$ and epoch in which the knot encounters the static shock $T_{1}$ for the new knots found in the VLBA-BU-BLAZAR data.}
    \begin{tabular}{cccc}
    \toprule
         & B9 & B10 & B11 \\
    \midrule
        $T_{0}$ (MJD) & $58378 \pm 51$  & $58813 \pm 26$ & $59535 \pm 115$ \\
        $T_{1}$ (MJD) & $58540\pm10$ & $59043\pm82$ & $59817\pm153$ \\
    \bottomrule
    
    \end{tabular}
    \label{tab:epochs}
\end{table*}

Moreover, Figure \ref{fig:0_separation} shows with dashed colored lines the period in which the new knots ``B9''-``B11'' passed the radio core, in blue, and passed the static shock ``A1'', in yellow. The green dashed line indicates the epoch when component ``B12'' was found. We followed the approach described in \cite{weaver_kinematics_2022} to estimate the ejection epoch, $T_{0}$, and the epoch when the knots encounter the static shock, $T_{1}$. Hence, for each component, we used either the first ten observed epochs or, in the case of accelerating knots, the points belonging to the first kinematic segment. Linear fits to the $x$ and $y$ coordinates were extrapolated back to the core position $(0,0)$ and to the location of the stationary shock ``A1''. This procedure provides the time of ejection along each coordinate, $T_{x_{0}}$ and $T_{y_{0}}$, and $T_{x_{1}}$ and $T_{y_{1}}$. Then the true time of ejection and its 1$\sigma$ uncertainty are given by:
\begin{equation}
\label{t0}
    T_{0} = \frac{T_{x_{0}}/\sigma_{T_{x_{0}}}^{2} + T_{y_{0}}/\sigma_{T_{y_{0}}}^{2}}{1/\sigma_{T_{x_{0}}}^{2} + 1/\sigma_{T_{y_{0}}}^{2}}
\end{equation}
\begin{equation}
\label{sigmat0}
    \sigma_{T_{0}} = \sqrt{\frac{(T_{0}-T_{x_{0}})^{2}/\sigma_{T_{x_{0}}}^{2} + (T_{0}-T_{y_{0}})^{2}/\sigma_{T_{y_{0}}}^{2}}{1/\sigma_{T_{x_{0}}}^{2} + 1/\sigma_{T_{y_{0}}}^{2}}}
\end{equation}
where $\sigma_{T_{x_{0}}}^{2}$ and $\sigma_{T_{y_{0}}}^{2}$ are the 1$\sigma$ propagated uncertainties on $T_{x_{0}}$ and $T_{y_{0}}$. The same equations were used to calculate $T_{1}$ and $\sigma_{T_{1}}$ substituting $T_{x_{0}}$, $\sigma_{T_{x_{0}}}$, $T_{y_{0}}$ and $\sigma_{T_{y_{0}}}$ with $T_{x_{1}}$, $\sigma_{T_{x_{1}}}$, $T_{y_{1}}$ and $\sigma_{T_{y_{1}}}$. 

Component ``B12'' is detected over too few epochs to allow a reliable calculation of either the epoch of ejection or its interaction with the stationary shock. Based on Figure \ref{fig:0_separation}, we argue that the ejection of ``B9'' and ``B10'', as well as their encounter with the static shock could be responsible for the flaring activity that occurred from 2019 to the beginning of 2020. Since the knot ``B9'' could be consistent with Component 24 of the \ac{MOJAVE} data, we calculated $T_{0}$ and $T_{1}$ using equations \ref{t0} and \ref{sigmat0} also for Component 24. In this case, we considered component 20 as the first static shock encountered by the knot since its distance from the core and its trend are consistent with ``A1''. The results are reported in Table~\ref{tab:my_label}. Similarly, the ejection and the shock interaction component ``B11'' could be responsible for the flaring activity from late 2021 through 2022. Hence, component ``B12'' and additional components that are still emerging could be responsible for all the flaring emissions from late 2023 to early 2024. In summary, we suggest that Flare 1 is associated with the interaction of ``B11'' with the standing shock, while Flare 2 to 7 could be linked to ``B12'', for which more VLBA observations are needed. However, the timing suggests the flaring began when the knot was still within the core or even before it reached the core. This and the appearance of further new VLBA components will be analyzed in Jorstad et al. {\it in prep.}

\begin{table}
    \centering
    \caption{Epoch of ejection $T_{0}$ and epoch in which the knot encounters the static shock $T_{1}$ for component 24.}
    \begin{tabular}{ccc}
    \toprule
        ID &$T_{0}$ (MJD) & $T_{1}$ (MJD)\\
        \midrule
        24 & $57586 \pm 233$ & $58673 \pm 450$\\
    \bottomrule
    \end{tabular}
    \label{tab:my_label}
\end{table}
 
\subsection{SED modeling}
\label{SED}
Section \ref{Radio_more} suggests a slightly different site for the flaring region within the jet for Flare 1 with respect to the flares from 2 to 7. Among these, the best sampled SED is the Flare 5 one, which is why we selected it for our Compton dominance analysis (see Section \ref{Compton Dominance}). In this section, we investigate whether we can model the spectral energy distributions of these two states.

For the modeling and fitting of the two radiative states of \op, we have used the JetSeT framework  \citep{tramacere_swift_2009, tramacere_stochastic_2011, tramacere_jetset_2020}. This is an open-source framework written in C/Python designed to simulate the radiative, accelerative processes, and adiabatic expansion occurring in relativistic jets and galactic objects, both beamed and unbeamed. This enables fitting numerical models to observational data, supporting the definition of complex numerical radiative scenarios, including synchrotron, \ac{SSC}, and \ac{EC} processes. The broadband emission has been interpreted in the context of a one-zone (blob) leptonic scenario, where the low-energy emission arises from synchrotron radiation, and the high-energy emission is interpreted as \ac{SSC}/\ac{EC} emission. The jet has conical geometry, with a semi-opening angle, $\theta_{\rm open}$. The emitting region has a spherical geometry with a radius $R$, a tangled magnetic field $B$, and it is moving with a bulk Lorentz factor $\Gamma$ along the jet axis, with a viewing angle $\theta$. The radius of the blob is implemented as a functionally dependent parameter in JetSeT, depending on its distance from the black hole (BH), $R_H$, and the jet opening angle according to: $R=R_H \tan~\theta_{\rm open}$. 

The electron population is characterized by a broken power-law distribution depending on the Lorentz factor, according to: 
\begin{equation}
n(\gamma) = 
\begin{cases} 
N K \gamma^{-p_1}, & \gamma_{\rm min} \leq \gamma < \gamma_{\rm break}, \\
N K \gamma_{\rm break}^{p_1 - p_{2}} \gamma^{-p_{2}}, & \gamma_{\rm break} \leq \gamma \leq \gamma_{\rm max}.
\end{cases}
\label{Eq:epopulation_bkn}
\end{equation}

where $\gamma_{\rm min}$, $\gamma_{\rm break}$, and $\gamma_{\rm max}$ are, respectively, the minimum, break, and maximum Lorentz factors, $p_1$ and $p_{2}$ are the spectral indices below and above the break, respectively, and $N$ is the number density of the emitters. The normalization factor $K$ is defined as
\begin{equation}
N = \int_{\gamma_{\rm min}}^{\gamma_{\rm max}} n(\gamma) \, d\gamma \, .
\end{equation}
In addition to the \ac{SSC} emission, we also considered the presence of external radiative fields emitted by the accretion disk, \ac{DT}, and the \ac{BLR}, contributing to the \ac{EC} emission. Furthermore, the \ac{EBL} model from \cite{franceschini_extragalactic_2008} was considered due to OP 313's high redshift.

The disk is modeled as a single-temperature blackbody, with the disk luminosity $L_{\rm Disk}$ and temperature $T_{\rm Disk}$. The value of temperature is set to $T_{disk}=\unit[5\times10^{4}]{K}$, while the luminosity is set at $L_{\rm Disk} = 5 \times 10^{45} \, \mathrm{erg/s}$ and frozen during the fit. This value has been chosen based on \cite{ghisellini_fermilat_2015}.

We assume that the DT reprocesses a fraction of $\tau_\text{DT}=0.1$ of the disk luminosity in the form of a black-body spectrum with temperature $T_\text{DT}=10^3\,$K. In addition, the radii of the \ac{BLR} and \ac{DT} are determined based on the disk luminosity according to \cite{kaspi_reverberation_2007} and \cite{cleary_spitzer_2007}, 
and are implemented as functionally-dependent parameters in JetSeT:
\begin{equation}
R_{BLR,in}=10^{17}\left(\frac{L_{\rm disc}}{10^{45}}\right)^{0.5} \mbox{cm}
\label{Eq:RBLRin}
\end{equation}
\begin{equation}
R_{\rm BLR,out}=1.1\times R_{\rm BLR,in}
\label{Eq:RBLRout}
\end{equation}
\begin{equation}
R_{\rm DT}=2.5\times10^{18}\left(\frac{L_{\rm disc}}{10^{45}}\right)^{0.5} \mbox{cm} \, 
\label{Eq:RDT}
\end{equation}

\begin{figure*}[htb]
    \sidecaption
    \includegraphics[width=12cm]{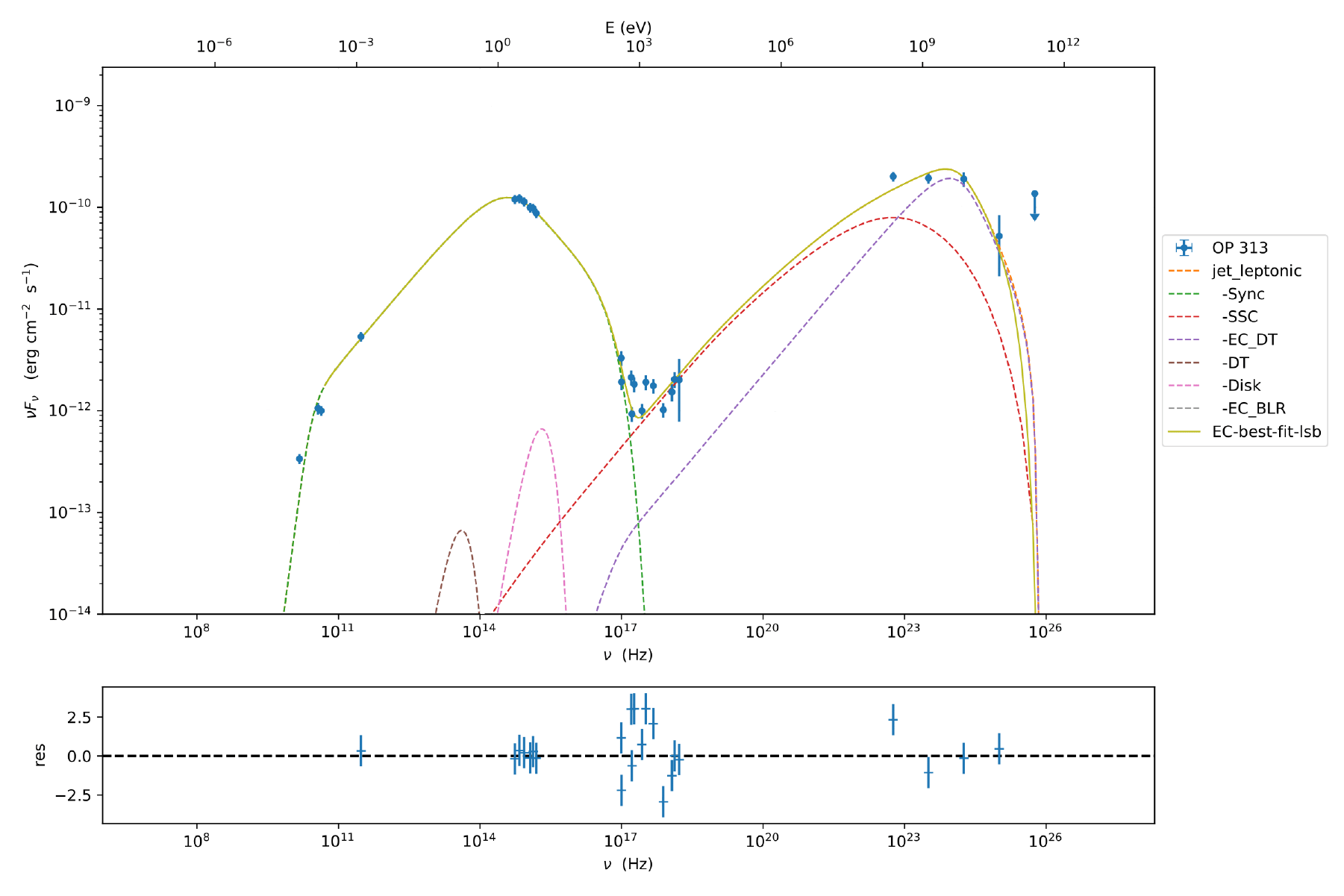}
    \caption{Best-fit model for the broadband \ac{SED} of OP 313 Flare 1 computed using the \texttt{JeTSet ModelMinimizer} module. The legend provides the color coding for the different components.}
    \label{fig:model1}
 \end{figure*}

\begin{figure*}[htb]
    \sidecaption
    \includegraphics[width=12cm]
    {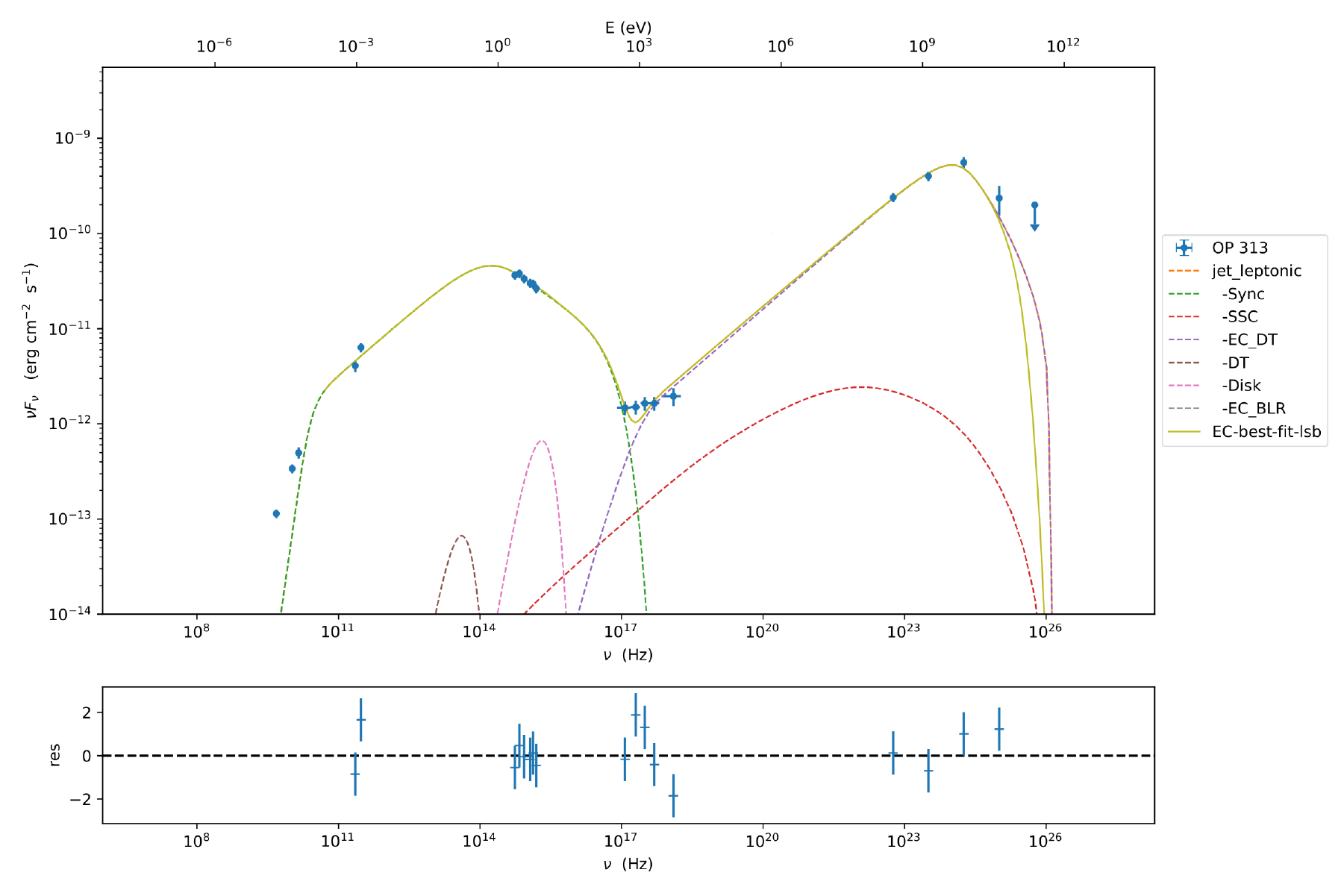}
    \caption{Best-fit model for the broadband \ac{SED} of OP 313 Flare 5, computed using the \texttt{JeTSet ModelMinimizer} module. The legend provides the color coding for the different components.}
    \label{fig:model5}         
\end{figure*}

We then performed a model fitting using the \texttt{JeTSet ModelMinimizer} module, which is based on the \texttt{iminuit} package \citep{dembinski_scikit-hepiminuit_2020}. The fit is performed using the data in Figure \ref{fig:SEDs} for both flaring periods, adding a systematic error of $10\%$ to all frequency ranges. Figure \ref{fig:model1} and Figure \ref{fig:model5} show the results of this fitting procedure. Furthermore, starting from the best-fit models found applying the \texttt{ModelMinimizer} module, we used the \texttt{JetSet McmcSampler} interface to the \texttt{emcee}\citep{foreman-mackey_emcee_2013} Python library to perform a Monte-Carlo-Markov chain (MCMC). Our goal was to obtain the posteriors of Flare 1 and Flare 5. The two resulting models with their confidence levels are shown in Figure \ref{fig:model_mcmc}.

Table \ref{ref:tableSED} shows the best-fit parameters, along with the corresponding uncertainties, resulting from the MCMC fit.

\begin{table}[htbp]
\centering
\caption{Posterior parameters for the broadband \acp{SED} of OP 313 in the Flare 1 and Flare 5.}

\begin{tabularx}{\columnwidth}{p{2.0cm} X X p{1cm}}
\toprule
Parameter                                                           & Flare 1                           & Flare 5                          & Frozen \\
\midrule

$\gamma_{min}$                                                      & $2.3^{+0.1}_{-0.1}$                   & $2.1^{+0.1}_{-0.1}$  & False     \\
$\gamma_{max}$                                                      & $(6.8^{+0.7}_{-0.8}) \times 10^4$    & $(9.4^{+0.4}_{-0.6})\times 10^{4}$               & False    \\
$N$ [cm$^{-3}$]                                                                 & $(1.5^{+0.3}_{-0.3}) \times 10^{1}$                   & $(2.0^{+0.5}_{-0.5})\times 10^{1}$  & False  \\
$\gamma_{break}$                                                    & $(6.9^{+1.3}_{-1.2})\times 10^3$      & $(5.8^{+1.0}_{-1.0})\times 10^3$ & False   \\
$p_1$                                                               & $2.0^{+0.1}_{-0.1}$ & $2.1^{+0.1}_{-0.1}$    & False  \\
$p_2$                                                               & $3.4^{+0.3}_{-0.3}$                   & $3.6^{+0.2}_{-0.2}$ & False  \\
*$R_{\rm DT}(L_{\rm Disk})$ [cm]                                         & $1.4\times 10^{19}$              & $1.4\times 10^{19}$              & True   \\
$\tau_{\rm DT}$                                                     & $0.1$                             & $0.1$                             & True   \\
$\tau_{\rm BLR}$                                                    & $0.1$                             & $0.1$                             & True   \\
$L_{Disk}$ [erg s$^{-1}$]                                                          & $5.0\times 10^{45}$               & $5.0\times 10^{45}$               & True     \\
$T_{ Disk}$ [K]                                                        & $5\times 10^4$     & $5\times 10^4$               & True    \\
*$R(R_H,\theta_{\rm open}$) [cm]                                         & $8.2\times 10^{17}$              & $8.78\times 10^{17}$              & True    \\
$R_H$ [cm]                                                               & $(8.1 ^{+1.4}_{-1.2})\times 10^{18}$ & $(9.1^{+1.9}_{-1.5})\times 10^{18}$   & False   \\
$B$ [G]                                                                & $(1.7^{+0.2}_{-0.1})\times 10^{-1}$  & $(8.5^{+0.8}_{-0.9})\times 10^{-2}$ & False   \\
$\theta$ [$^{\circ}$]                                                           & $3.5$                   & $3.5$
 & True    \\
$\theta_{\rm open}$ [$^{\circ}$]                                                & 5.00                       &  5.00                      & True    \\
$\Gamma$                                                            & $13.1^{+0.2}_{-0.2}$                  & $14.6^{+0.2}_{-0.2}$  & False  \\
\bottomrule
\end{tabularx}
   \label{ref:tableSED}
   \tablefoot{ $\gamma_{\rm min}$, $\gamma_{\rm break}$, and $\gamma_{\rm max}$ are, respectively, the minimum, break, and maximum Lorentz factors, $p_1$ and $p_{2}$ are the spectral indices below and above the break, respectively, N is the emitter density, $R_{\rm DT}$ is the dusty torus radius, $\tau_{\rm DT}$ and $\tau_{\rm BLR}$ are, respectively, the \ac{BLR} and \ac{DT} optical depth, $R_{\rm BLR,in}$ and $R_{\rm BLR,out}$ are the inner and outer \ac{BLR} radius, $L_{disk}$ is the disk luminosity, $T_{disk}$ is the temperature of the disk, R is the emitting region with a spherical geometry radius, $R_H$ is the emitting region position in the jet, B is the tangled magnetic field, $\theta$ is the viewing angle, $\theta_{\rm open}$ is the semi-opening angle of the conical jet, $\Gamma$ is the bulk Lorentz factor along the jet axis.}
\end{table}
 
We computed the magnetic energy density over the electron energy density ($u_{B}$/$u_{e}$) ratio for the two flaring periods to assess the magnetization of the emitting region. Using the values reported in Table \ref{ref:tableSED}, we obtained $u_{B}$/$u_{e}$ values equal to 0.031 and  0.013 for Flare 1 and Flare 5, respectively, with 
\begin{equation}
u_{B}=B^{2}/8\pi    
\end{equation}
\begin{equation}
u_e = m_e c^2 \int_{\gamma_{\min}}^{\gamma_{\max}} \gamma \, n(\gamma) \, d\gamma
\end{equation}
where $m_{e}c^{2}$ is the rest energy of the electron and $n(\gamma)$ is the electron population described in Equation \ref{Eq:epopulation_bkn}. These values indicate that the emission region is particle-dominated, with a relatively low magnetic energy density compared to the energy carried by relativistic electrons. 

The two-flaring period \acp{SED} exhibits the typical two-bump structure commonly associated with blazars, where the low-energy peak is attributed to synchrotron radiation and the high-energy peak arises from inverse Compton scattering. However, the relative contributions of different components vary between the two flaring periods, reflecting distinct physical conditions in the jet and its surroundings. The observed fluxes for both flaring periods at high energy are well explained by the model with \ac{SSC} and \ac{EC} emission mechanisms, as seen in Figure \ref{fig:model1} and \ref{fig:model5}. In both cases, the \ac{EC} dominates over the \ac{SSC} emission,  with a lower \ac{CD} in the first one ($\rm CD\approx 1$). 

As discussed in \ref{Radio_more}, the $\gamma$-ray flares are likely associated with new components merging from the VLBA core, which places the emission regions rather far from the central engine. This is also in agreement with the studies of \cite{kramarenko_decade_2022} and \cite{costamante_origin_2018} that places the $\gamma$-ray emission region outside the \ac{BLR}, and with the detection of the source up to energies 250\,GeV by LST-1 and MAGIC \citep{takeishi_pushing_2025}. If the emission region was inside the \ac{BLR}, the $\gamma$-rays of such high energies would be absorbed \citep[e.g.][]{tavecchio_intrinsic_2009}.

Hence, for both flaring periods, we set $R_{H}$ to have the emission region outside the \ac{BLR} and show the \ac{DT} as the main source of seed photons. The change in the parameters between Flare 1 and 5, i.e. an increase of $R_H$, and a consequent increase of $R$ (due to the jet geometry), with a decrease of the magnetic field, are qualitatively consistent with an adiabatically expanding blob moving along the axis of a conical jet. This could be in agreement with the results presented in Section \ref{Radio_more}. This scenario, investigated in \cite{tramacere_radio-_2022}, would also imply a very mild decrease in the total number of particles (if particle escape is very inefficient), and a magnetic field value dictated by the flux-freezing theorem $B(R(t))=B_0(R_0/R(t))^{m_B}$ \citep{begelman_theory_1984}. However, this interpretation is not unique. The inferred distance traveled in time between flares suggests only mildly relativistic motion, which is inconsistent with the assumed bulk Lorentz factor. A purely adiabatic expansion would likely produce a more continuous high-flux state rather than temporally distinct flares, as we see in the lightcurve. As shown in Section \ref{Radio_more}, we found that before the 2024 flaring activity, new components arose from the jet and could collide with the first static shock. Hence, it is likely that an additional shock acceleration occurred during Flare 5, on top of the expansion. Hence, while our modeling suggests the adiabatic expansion as a plausible scenario, alternative combinations of parameters (e.g., lower $B$ or higher $\Gamma$) and physical mechanisms, like turbulence and kink instabilities, may yield comparable fits. Further multiwavelength constraints would be required to break this degeneracy. In particular, the obtained magnetic fields and Doppler factors are consistent with internal shocks within the relativistic jets \citep[e.g.,][]{marscher_models_1985, bottcher_timing_2010}, but moderate changes of these parameters could also point to turbulence-driven particle acceleration \citep[e.g.,][]{marscher_turbulent_2014, chen_particle_2016} or magnetic reconnection and kink-instability scenarios \citep[e.g.,][]{giannios_fast_2009, barniol_duran_simulations_2017, tchekhovskoy_three-dimensional_2016}.

\cite{dar_signature_2025} proposes that a hadronic component contributed to the flaring emission of OP 313.\cite{dar_signature_2025} suggests that, in the period in which we identify the Flare 2 and 3, the inclusion of the photo-meson process along with the leptonic emission component can successfully reproduce the \ac{SED}, while the leptonic one cannot. However, the optical-gamma correlation we see suggests a rather leptonic origin (see Section \ref{correlation}).
We also note that this is the period in which the \ac{VHE} gamma-ray emission from \op\, was detected, and the goodness of the leptonic fit will be further investigated in the upcoming paper by MAGIC and LST Collaborations \citep{takeishi_pushing_2025}. 
 
\section{Summary}
\label{Conclusion}

We have presented multiwavelength observations of the high redshift \ac{FSRQ} \op \, during $2008$ - $2024$, including radio-to-$\gamma$-ray data collected by \ac{Fermi-LAT}, \ac{Swift}, ATLAS, CRTS, KAIT, Tuorla, ZTF, Metshäovi, VLBA, QUIVER, SMAPOL, and the F-GAMMA project. Strong variability has been observed from optical to $\gamma$ rays, with an increasing flux starting from 2022. The optical, UV, X-ray, and $\gamma$-ray light curves are well correlated. A different behaviour has been observed in radio, with the flux being high in 2008 and a new increasing trend in 2019, without a clear correlation with the other bands. 

Focusing on the $\gamma$-ray observations, in particular those after 2021, we identified 8 flaring periods and investigated the interplay between particle acceleration and radiative cooling in the source's jet by means of hysteresis plots. We found no clear hysteresis pattern for the first, second, and fifth flaring periods. The \ac{SED} of the fifth $\gamma$-ray flaring period shows a significant Compton dominance, suggesting that photon fields outside the jet are responsible for the IC scattering. This is in agreement with the results shown in \cite{kramarenko_decade_2022} that identify the $\gamma$ -ray emission region to be outside the \ac{BLR}. Moreover, an overall correlation has been observed between optical and $\gamma$ -ray bands without significant time lag, suggesting that the emission in these two bands has a common origin and is produced by leptonic processes. 

 The kinematics of the jet were investigated through the analysis of high-resolution radio images collected from the VLBA-BU-BLAZAR and \ac{MOJAVE} projects. New components were found to arise in the jet starting from 2021, and we argued that some of them could be responsible for the flaring activity observed since 2022. To investigate this possibility, we calculated the epoch of ejection and the epoch at which the knots encountered a standing shock in the jet. As a result, we claim that the knots named ``B11'' could be responsible for the flaring activity observed in 2022-2023. We modeled the \acp{SED} of the first and fifth flaring periods using a single-zone model that includes both \ac{SSC} and \ac{EC} components to test the scenario in which the $\gamma$-ray emitting region is located far from the central engine and the dusty torus acts as the dominant source of seed photons. This analysis shows that the \ac{DT} component is more prominent in the fifth flaring period, in agreement with expectations. This highlights the crucial role of the \ac{DT}'s photon field in accounting for the $\gamma$-ray emission of \op.
 This outcome is also consistent with the detection of the source at energies up to $250\,\mathrm{GeV}$ by the LST-1 and MAGIC telescopes, since photons of such high energy would be absorbed if emitted within the BLR \citep[e.g.][]{tavecchio_intrinsic_2009}. Therefore, detailed modeling of the spectral energy distribution during flare 3, using the VHE $\gamma$-ray data (LST-1 and MAGIC Collaborations, {\it in prep.}), could further strengthen this conclusion. Further study of jet kinematics, including the most recent VLBA data (Jorstad et al., {\it in prep.}), will also be valuable in confirming the ‘far-from-the-black-hole’ emission scenario outlined in this paper.
 
\section*{Acknowledgments}
\begin{acknowledgements}
Author Contributions.  C. Bartolini: project leadership, Fermi-LAT data analysis, Swift data analysis, \ac{VLBA} data analysis, theoretical modeling, paper drafting and editing; E. Lindfors: optical data analysis, interpretation, and paper editing; D. Cerasole: Swift data analysis, optical and $\gamma$-ray correlation study, paper editing; A. Tramacere: theoretical modeling and paper drafting; A. L\"ahteenm\"aki and M. Tornikoski: Mets$\ddot{a}$hovi data analysis. The rest of the authors have contributed in one or several of the following ways: data acquisition, processing, calibration, and/or reduction; cross-check of the data analysis; draft editing.

C.B.: This paper and related research have been conducted during and with the support of the Italian national inter-university PhD program in Space Science and Technology. 
C.B.: Thanks to Beyoncè's songs for giving her the necessary strength to write this paper during a difficult time.
F.G.: acknowledges financial support from Junta de Castilla y León
project  SA101P24.
J.J.: was supported by Academy of Finland projects 320085 and 345899
P.K.: was supported by Academy of Finland projects 346071 and 345899.

The \textit{Fermi}-LAT Collaboration acknowledges generous ongoing support from a number of agencies and institutes that have supported both the development and the operation of the LAT as well as scientific data analysis. These include the National Aeronautics and Space Administration and the Department of Energy in the United States, the Commissariat \`a l'Energie Atomique and the Centre National de la Recherche Scientifique / Institut National de Physique Nucl\'eaire et de Physique des Particules in France, the Agenzia Spaziale Italiana and the Istituto Nazionale di Fisica Nucleare in Italy, the Ministry of Education, Culture, Sports, Science and Technology (MEXT), High Energy Accelerator Research Organization (KEK) and Japan Aerospace Exploration Agency (JAXA) in Japan, and the K.~A.~Wallenberg Foundation, the Swedish Research Council and the Swedish National Space Board in Sweden.

Additional support for science analysis during the operations phase is gratefully acknowledged from the Istituto Nazionale di Astrofisica in Italy and the Centre National d'\'Etudes Spatiales in France. This work performed in part under DOE Contract DE-AC02-76SF00515.

This research has made use of the NASA/IPAC Infrared Science Archive, which is funded by the National Aeronautics and Space Administration and operated by the California Institute of Technology.

This publication makes use of data obtained at Mets\"ahovi Radio Observatory, operated by Aalto University in Finland. 

This publication makes use of data based on observations with the $100\, \mathrm{m}$ telescope of the MPIfR (Max-Planck-Institut für Radioastronomie). I.M., I.N. and V.K. were funded by the International Max Planck Research School (IMPRS) for Astronomy and Astrophysics at the Universities of Bonn and Cologne.

This research has made use of data from the MOJAVE database that is maintained by the MOJAVE team \citep{lister_mojave_2018}.

This study makes use of VLBA data from the VLBA-BU Blazar Monitoring Program (BEAM-ME and VLBA-BU-BLAZAR; \url{http://www.bu.edu/blazars/BEAM-ME.html}), funded by NASA through the Fermi Guest Investigator Program. The VLBA is an instrument of the National Radio Astronomy Observatory. The National Radio Astronomy Observatory is a facility of the National Science Foundation operated by Associated Universities, Inc.

This research has made use of data from the MOJAVE database that is maintained by the MOJAVE team \citep{lister_mojave_2018}.

Partly based on observations with the 100-m telescope of the MPIfR (Max-Planck-Institut f\"ur Radioastronomie) at Effelsberg. Observations with the 100-m radio telescope at Effelsberg have received funding from the European Union's Horizon 2020 research and innovation programme under grant agreement No 101004719 (ORP).

The Submillimeter Array is a joint project between the Smithsonian Astrophysical Observatory and the Academia Sinica Institute of Astronomy and Astrophysics and is funded by the Smithsonian Institution and the Academia Sinica. Maunakea, the location of the SMA, is a culturally important site for the indigenous Hawaiian people; we are privileged to study the cosmos from its summit.

This work has made use of data from the Joan Oró Telescope (TJO) of the Montsec Observatory (OdM), which is owned by the Catalan Government and operated by the Institute for Space Studies of Catalonia (IEEC). Based on observations obtained with the Samuel Oschin Telescope 48-inch and the 60-inch Telescope at the Palomar Observatory as part of the Zwicky Transient Facility project. ZTF is supported by the National Science Foundation under Grant No. AST-2034437 and a collaboration including Caltech, IPAC, the Weizmann Institute for Science, the Oskar Klein Center at Stockholm University, the University of Maryland, Deutsches Elektronen-Synchrotron and Humboldt University, the TANGO Consortium of Taiwan, the University of Wisconsin at Milwaukee, Trinity College Dublin, Lawrence Livermore National Laboratories, and IN2P3, France. Operations are conducted by COO, IPAC, and UW. The ZTF forced-photometry service was funded under the Heising-Simons Foundation grant \#12540303 (PI: M.J.Graham). This work has made use of data from the Asteroid Terrestrialimpact Last Alert System (ATLAS) project. The Asteroid Terrestrial-impact Last Alert System (ATLAS) project is primarily funded to search for near earth asteroids through NASA grants NN12AR55G, 80NSSC18K0284, and 80NSSC18K1575; byproducts of the NEO smarch include images and catalogs from the survey area. This work was partially funded by Kepler/K2 grant J1944/80NSSC19K0112 and HST GO-15889, and STFC grants ST/T000198/1 and ST/S006109/1. The ATLAS science products have been made possible through the contributions of the University of Hawaii Institute for Astronomy, the Queen’s University Belfast, the Space Telescope Science Institute, the South African Astronomical Observatory, and The Millennium Institute of Astrophysics (MAS), Chile.
    
\end{acknowledgements}

\bibliographystyle{aa}
\bibliography{references}{}

\begin{appendix} 
\section{Hysteresis patterns in 2-day long time bins}
\label{Appendix hysteresis}
\begin{figure*}[htb]
    \centering
    \subfigure[\label{fig:hysteresis1Flare2days}]{\includegraphics[width=0.3\linewidth]{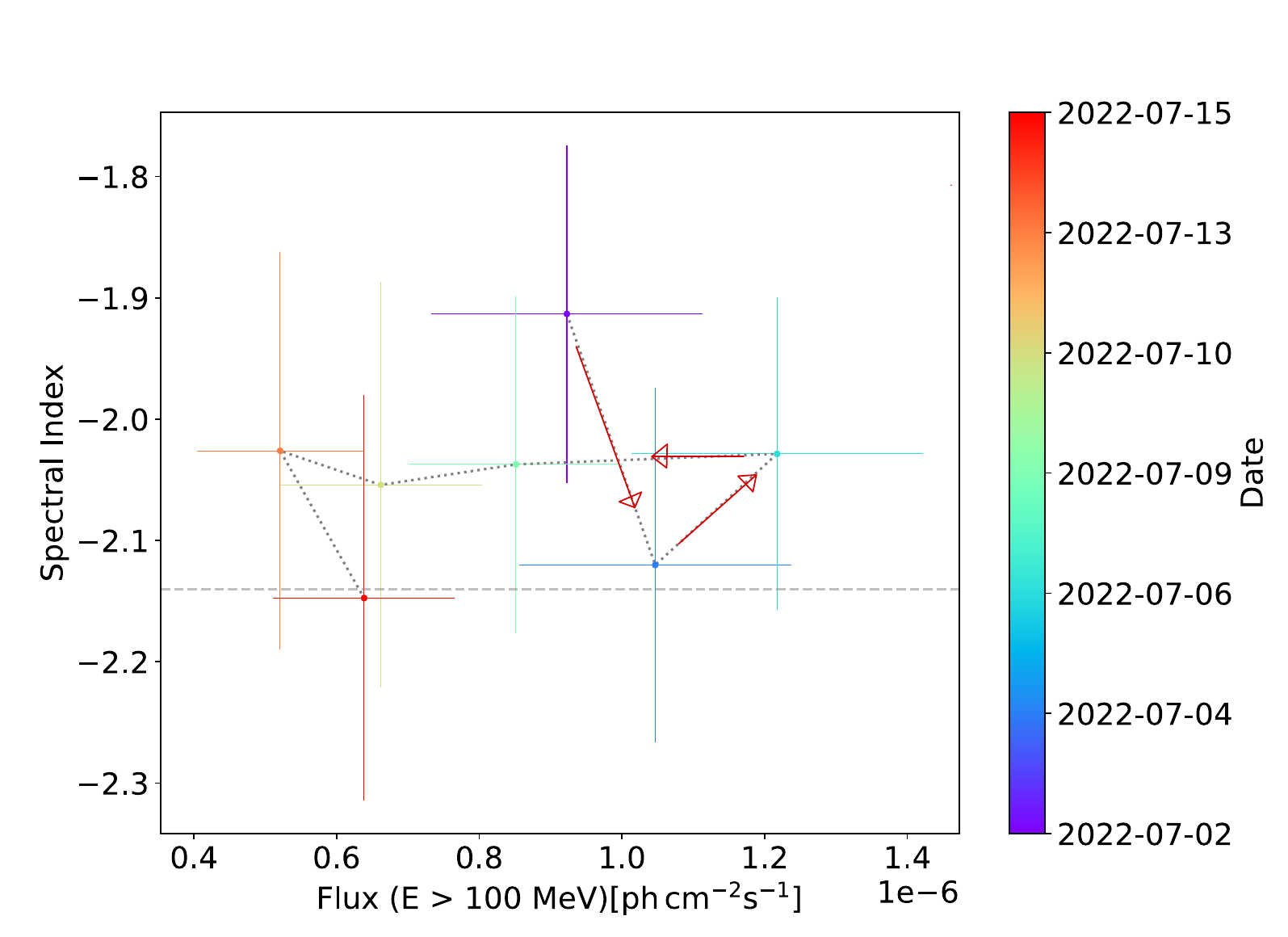}}\quad
    \subfigure[\label{fig:hysteresis5Flare2days}]{\includegraphics[width=0.30\linewidth]{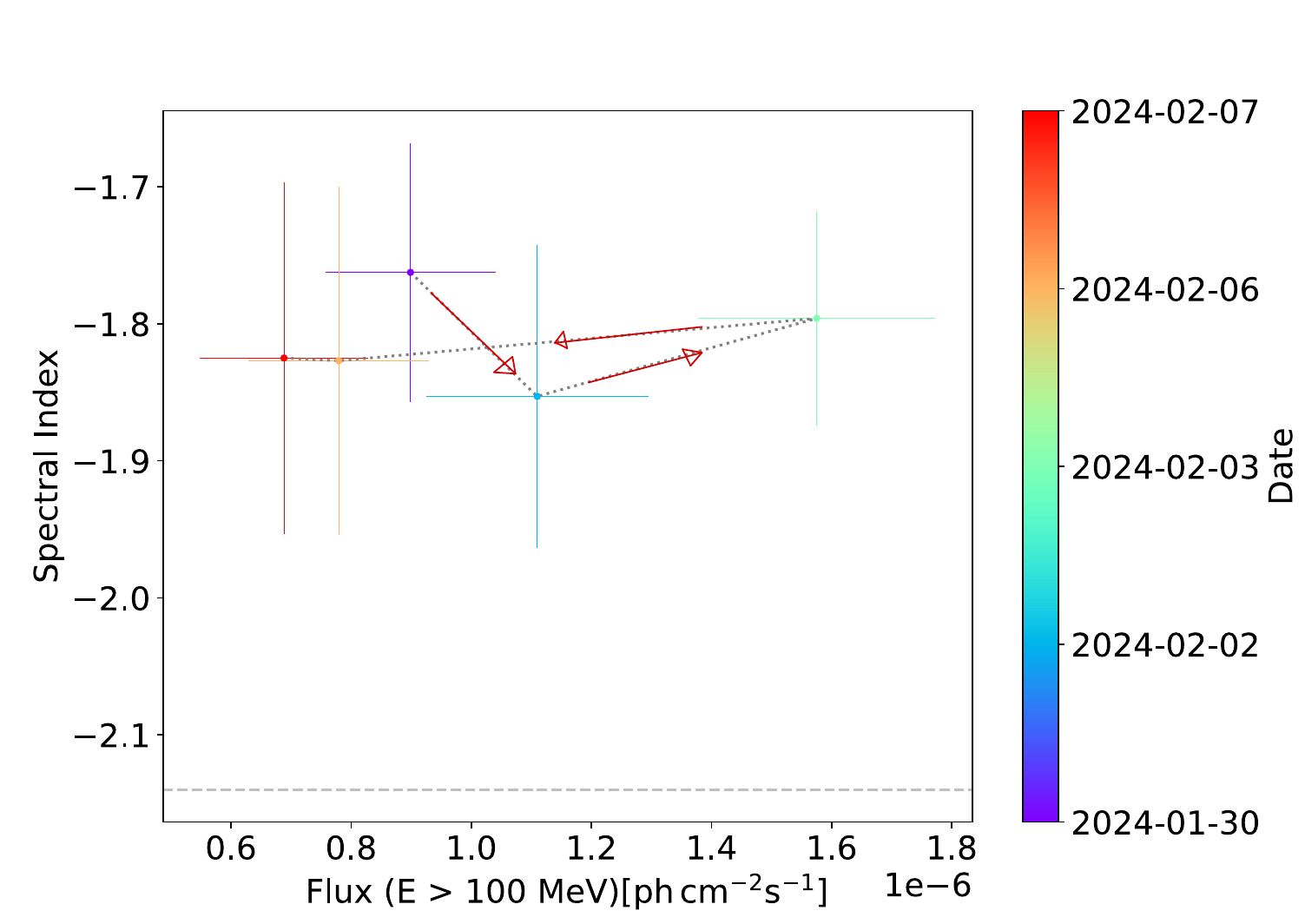}}\quad
    \subfigure[\label{fig:hysteresis2Flare2days}]{\includegraphics[width=0.36\linewidth]{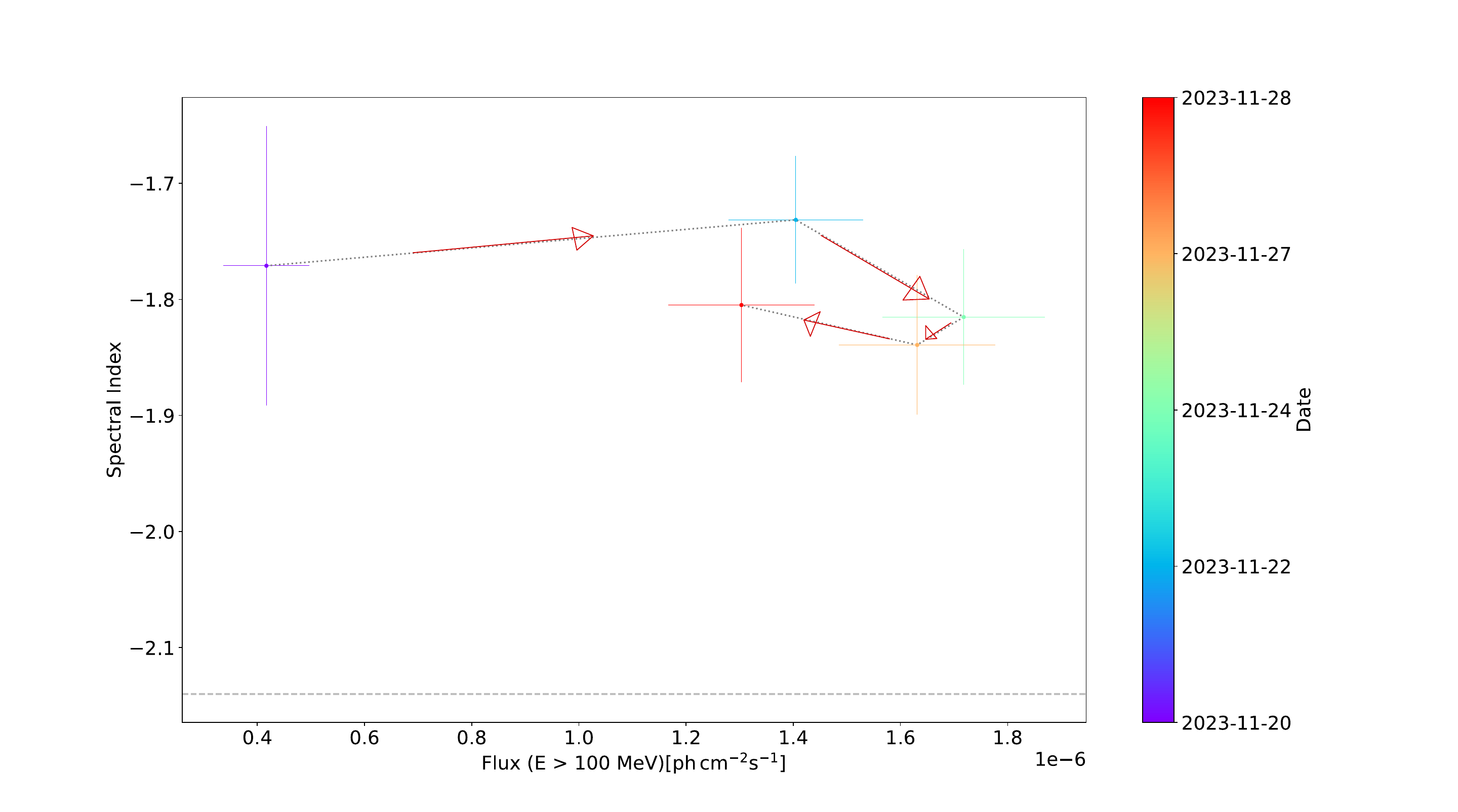}}
    \caption{a) First flaring period hysteresis pattern from the 2nd of July 2022 to the 15th of July 2022, considering 2-day-long time bins. The grey dashed line is drawn to help seek the correct order of the points. b) Fifth flaring period hysteresis pattern from the 30th of January 2024 to the 8th of February 2024, considering 2-day-long time bins. c) Second flaring period hysteresis pattern from the 20th of November 2023 to the 29th of November 2023, considering 2-day-long time bins.}    
\end{figure*}

In Figure \ref{fig:hysteresis1Flare2days}, \ref{fig:hysteresis5Flare2days}, and \ref{fig:hysteresis2Flare2days}, we show how the hysteresis patterns of the first, the second, and the fifth flaring periods are if we consider the average photon index in 2-day-long time bins. The hysteresis pattern's hints we find are anti-clockwise for the first and the fifth flaring periods, and clockwise for the second flaring period. This feature shows why we can not establish that our patterns are truly clockwise or anti-clockwise due to the large photon index error bars, and how they and the average flux change considering different time bins.

\section{MOJAVE and VLBA-BU-BLAZAR tables}
\label{Appendix radio}
Table \ref{tab:MOJAVE} describes the fits as follows: 1- ID of the component; 2— mean flux density and error in mJy; 3- mean projected distance from the core feature in mas; 4- mean position angle of feature with respect to the core component; 5- angular proper motion in milliarcseconds per year; 6- proper motion in units of the speed of light; 7- acceleration in milliarcsecond per square year; 8- date of reference (middle) epoch used for fit.

\begin{sidewaystable*}
\centering
\captionsetup{justification=centering}
\caption{Jet structure and kinematic parameters of the \ac{MOJAVE} dataset.}
\label{tab:MOJAVE}
\begin{tabular}{cccccccc}
\toprule
  & & & & MOJAVE & & &  \\
ID & <S> ($\unit{mJy}$) & <R> ($\unit{mas}$) & P.A. ($\unit{deg}$) & $\mu$ ($\unit{m as} / yr$) & $\beta_{app}$ & $\dot{\mu}$ ($\unit{mas} / yr^{2}$) & Middle epoch \\
\midrule
2 & $0.020\pm0.007$ & $8.94\pm0.19$& $284.6\pm15$& $0.345\pm0.039$ & $18.06\pm2.06$ & $0.018\pm0.008$& 2016.69\\
8 & $0.028\pm0.001$ & $1.57\pm0.19$& $300.3\pm21.4$& $0.272\pm0.048$ & $14.24\pm2.52$ & $0.210\pm0.09$& 2009.93\\
11 & $0.026\pm0.01$ & $5.54\pm0.18$& $302.1\pm5.5$& $0.436\pm0.007$ & $22.84\pm0.36$& $0.01\pm0.003$ & 2013.34\\
17 & $0.261\pm0.013$& $1.28\pm0.16$& $309.1\pm39$& $0.375\pm0.005$ & $19.64\pm0.27$& $0.029\pm0.003$& 2013.84\\
18 & $0.214\pm0.011$& $1.84\pm0.17$& $303.8\pm47.6$& $0.382\pm0.019$& $20.01\pm1.46$& $0.018\pm0.007$& 2016.51\\
19 & $0.057\pm0.003$& $1.06\pm0.19$& $284.6\pm16.2$& $0.193\pm0.028$ & $10.11\pm1.46$ &$0.015\pm0.014$ & 2018.44\\
20 & $0.516\pm0.026$& $0.24\pm0.18$& $295.2\pm6.7$& $0.007\pm0.002$& $0.16\pm0.03$& $0.002\pm 0.001$& 2018.42\\
22 & $0.013\pm0.001$& $3.57\pm0.17$& $294.2\pm37$& $0.157\pm0.032$& $8.23\pm1.71$& $0.026\pm0.01$&  2017.96 \\
24 & $0.223\pm0.011$& $0.54\pm0.16$& $316.1\pm33.7$& $0.085\pm0.011$ & $4.42\pm0.59$ &  & 2022.80\\
\bottomrule
\end{tabular}
\end{sidewaystable*}
Table \ref{tab:Boston} describes the fits as follows: 1- ID of the component; 2— mean flux density and error in mJy; 3- mean projected distance from the core feature in mas; 4- mean position angle of feature with respect to the core component; 5- angular proper motion in milliarcseconds per year; 6- proper motion in units of the speed of light; 7- parallel acceleration the jet in milliarcsecond per square year; 8- perpendicular acceleration the jet in milliarcsecond per square year.

\begin{sidewaystable*}
\centering
\captionsetup{justification=centering}
\caption{Jet structure and kinematic parameters of the VLBA-BU-BLAZAR dataset.}
\label{tab:Boston}
\begin{tabular}{cccccccc}
\toprule
& & & & VLBA-BU-BLAZAR Boston &  & & \\
ID & <S> ($\unit{Jy}$) & <R> ($\unit{mas}$) & P.A. ($\unit{deg}$) & $\mu$ ($\unit{m as} / yr$) & $\beta_{app}$ & $\dot{\mu}_{\parallel}$ ($\unit{mas} / yr^{2}$) & $\dot{\mu}_{\perp}$ ($\unit{mas} / yr^{2}$)  \\
\midrule
B1 & $0.09\pm0.01$  & $0.29 \pm 0.01$ & $-0.1 \pm 3.1$  &$0.100 \pm 0.007$ & $5.25 \pm 0.35$ & &  \\
B2($1^{st}$) & \multirow{2}{*}{$0.24\pm0.02$} & \multirow{2}{*}{$0.63 \pm 0.1$} & \multirow{2}{*}{$-50.1\pm 2$} & $0.118 \pm 0.003$ & $6.16 \pm 0.16$ & \multirow{2}{*}{$0.086\pm 0.015$} &  \multirow{2}{*}{$-0.018\pm 0.018$} \\
B2($2^{nd}$) &  &  &  & $0.294 \pm 0.029$ & $15.39 \pm 1.53$ & & \\
B3 & $0.4\pm0.02$ & $0.37 \pm 0.02$  & $-67.4\pm 1.5$ & $0.275 \pm 0.007$ & $14.38 \pm 0.36$ &  &   \\
B4($1^{st}$) & \multirow{2}{*}{$0.27\pm0.02$} & \multirow{2}{*}{$0.57 \pm 0.08$}  & \multirow{2}{*}{$-64.5\pm3.3$} & $0.128 \pm 0.004$ & $6.73 \pm 0.21$ & \multirow{2}{*}{$0.220\pm 0.006$} & \multirow{2}{*}{$-0.021\pm 0.007$} \\
B4($2^{nd}$) &  &   &  & $0.502 \pm 0.008$ & $26.29 \pm 0.40$ &  &  \\
B5 & $0.03\pm0.01$ & $1.12\pm 0.22$  & $-56.1\pm6.7$ & $0.611 \pm 0.014$ & $32.00\pm 0.74$ &  &   \\
B6 & $0.07\pm0.01$ & $0.67 \pm 0.11$ & $-83.9\pm 7.4$ & $0.271 \pm 0.017$& $14.16 \pm 0.87$ &  &   \\
B7 & $0.05\pm0.01$ & $0.35 \pm 0.03$ & $-77.6\pm3.9$ & $0.085 \pm 0.002$ & $4.46 \pm 0.11$ &  &   \\
B8 & $0.05\pm0.01$ & $0.31 \pm 0.05$ & $-72.4\pm8.2$ & $0.013 \pm 0.003$ & $0.68 \pm 0.17$ &  &   \\
B9 & $0.26\pm0.01$  & $0.43 \pm 0.02$ & $-45.1\pm9$ & $0.085 \pm 0.009$ & $4.46 \pm 0.46$ &  & \\
B10 & $0.24\pm0.01$  & $0.51 \pm 0.03$ & $-53.8\pm12.7$ & $0.123 \pm 0.007$ & $6.44 \pm 0.34$ &  &  \\
B11 & $0.21\pm0.01$  & $0.4 \pm 0.03$ & $-49.7\pm18$ & $0.175 \pm 0.007$ & $9.18 \pm 0.35$ & & \\
B12 & $0.58\pm0.01$  & $0.18 \pm 0.01$ & $-66.8\pm11.4$ & $0.033 \pm 0.003$ & $1.73 \pm 0.16$ & & \\
\bottomrule

\end{tabular}
\end{sidewaystable*}

\section{SED modeling table and plots}
\label{Appendix SED}
\begin{figure}[htb]
    \subfigure[\label{fig:model1m}]
        {\includegraphics[width=0.8\linewidth]{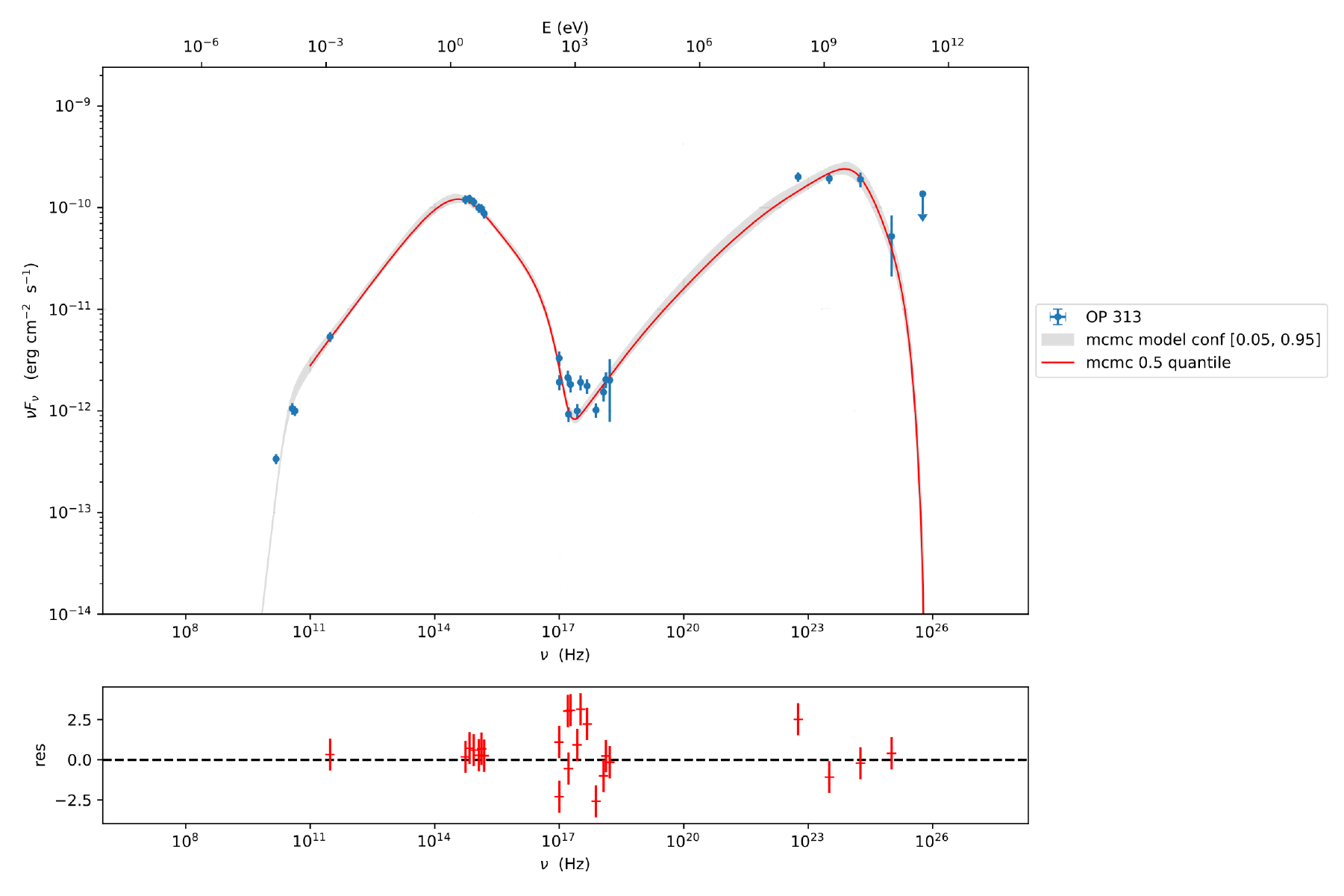}}\quad
        \subfigure[\label{fig:model5m}]        {\includegraphics[width=0.8\linewidth]{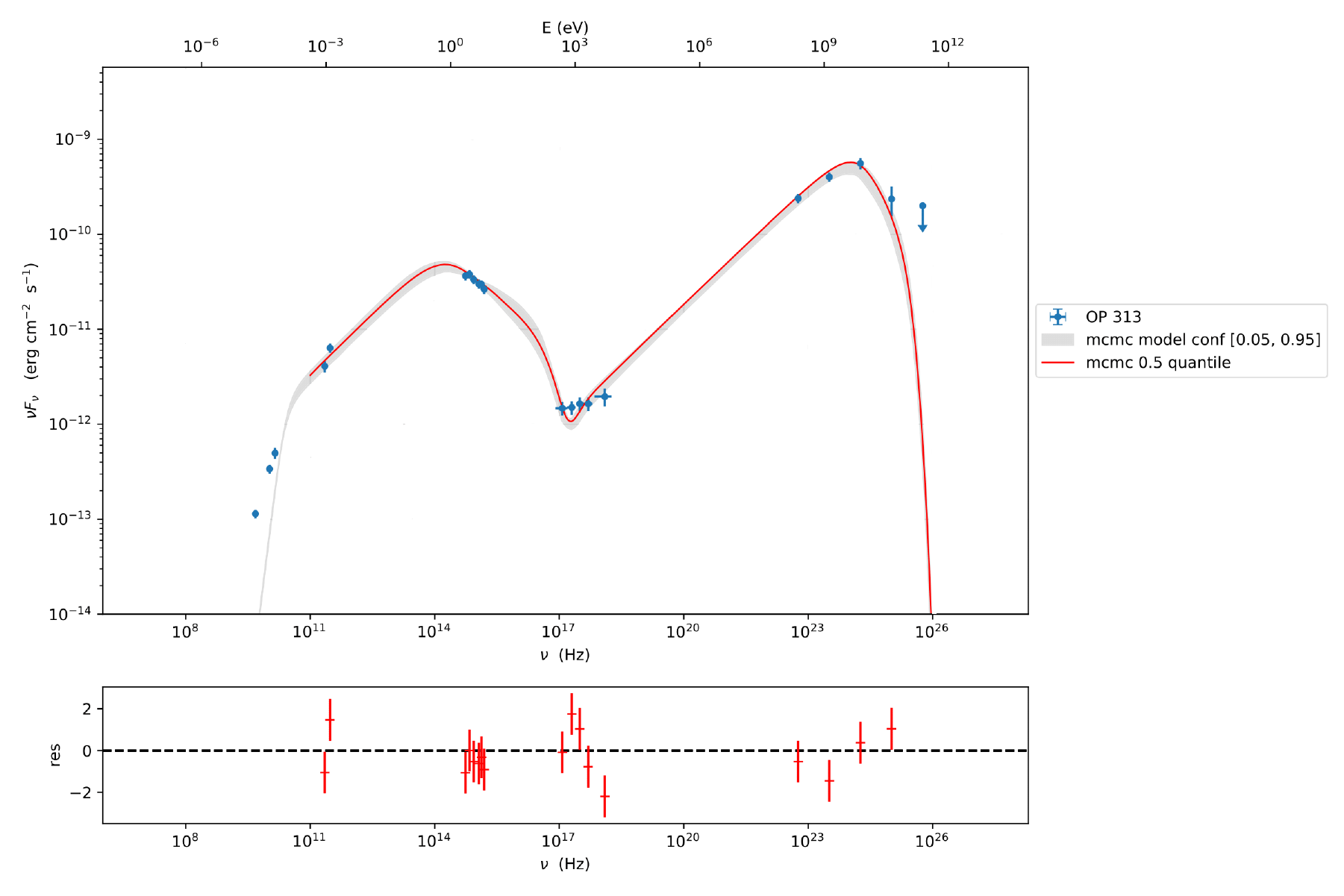}} 
    \caption{(a) Model-fit broadband \ac{SED} of OP 313's Flare 1 and (b) model-fit broadband \ac{SED} of OP 313's Flare 5 using \texttt{JeTSet McmcSampler}. The legends provide the colour coding for the MCMC information.}
    \label{fig:model_mcmc}
\end{figure}

Figure \ref{fig:model1m} shows the broadband \ac{SED} of OP 313's Flare 1 and \ref{fig:model5m} shows the broadband \ac{SED} of OP 313's Flare 5 using \texttt{JeTSet McmcSampler}.

\end{appendix}

\end{document}